\documentclass[3p,times]{elsarticle}

\usepackage[latin9]{inputenc}
\usepackage{verbatim}
\usepackage{lmodern}
\usepackage{graphicx}
\usepackage{amsmath,mathrsfs,amssymb}
\usepackage{amsfonts}
\usepackage{caption}
\usepackage{array}
\usepackage{longtable}
\usepackage{algpseudocode}
\usepackage{amsmath}
\usepackage{mathpazo}
\usepackage{float}
\usepackage{algorithm}
\graphicspath{ {images/} }
\usepackage{epstopdf}
\usepackage{enumitem}
\usepackage{booktabs}
\usepackage{subcaption}
\usepackage{lineno}
\usepackage[english]{babel}
\usepackage[unicode=true,
 bookmarks=false,
 breaklinks=false,pdfborder={0 0 1},colorlinks=false]
 {hyperref}
\hypersetup{
 colorlinks,bookmarksopen,bookmarksnumbered,citecolor=red,urlcolor=red}

\makeatletter

\pdfpageheight\paperheight
\pdfpagewidth\paperwidth

\modulolinenumbers[5]

\newtheorem{defn}{Definition}

\newtheorem{lem}{Lemma}

\newtheorem{thm}{Theorem}
\newtheorem{rem}{Remark}

\newtheorem{assum}{Assumption}

\begin{document}

\noindent\rule{16.4cm}{2pt}\\
\underline{To cite this article:}
{\bf{\textcolor{red}{H. A. Hashim, L. J. Brown, and K. McIsaac, "Nonlinear Stochastic Position and Attitude Filter on the Special Euclidean Group 3," Journal of the Franklin Institute, vol. 356, no. 7, pp. 4144-4173, 2019.}}}\\
\noindent\rule{16.4cm}{2pt}\\

\noindent{\bf The published version (DOI) can be found at:  \href{https://doi.org/10.1016/j.jfranklin.2018.12.025}{10.1016/j.jfranklin.2018.12.025} }\\

\footnotesize{
\begin{center}
	\vspace{40pt}
	\noindent Please note that where the full-text provided is \underline{\bf the preprint version} and this may differ from the revised and/or the final Published\\
\end{center}
\begin{center}
    version. {\bf To cite this publication, please use the final published version.}  \vspace{10pt}
\end{center}

	\textbf{
		\begin{center}
			\vspace{20pt}
			Personal use of this material is permitted. Permission from the author(s) and/or copyright holder(s), must be obtained for all other uses, in any current or future media, including reprinting or republishing this material for advertising or promotional purposes.\\
		\end{center}
	\vspace{335pt}
		\vspace{20pt}Please contact us and provide details if you believe this document breaches copyrights. We will remove access to the work immediately and investigate your claim.
} 
}

\normalsize
\begin{frontmatter}

\title{Nonlinear Stochastic Position and Attitude Filter on the Special Euclidean Group 3\tnoteref{mytitlenote}}

\author[HAH]{Hashim A.~Hashim\corref{cor1}}
\ead{hmoham33@uwo.ca}
\author[LJB]{Lyndon J. Brown}
\ead{lbrown@uwo.ca}
\author[KM]{Kenneth McIsaac}
\ead{kmcisaac@uwo.ca}

\address[HAH,LJB,KM]{Department of Electrical and Computer Engineering, University of Western Ontario, London, ON, N6A-5B9, Canada}

\begin{abstract}
This paper formulates the pose (attitude and position) estimation
problem as nonlinear stochastic filter kinematics evolved directly
on the Special Euclidean Group $\mathbb{SE}\left(3\right)$. This
work proposes an alternate way of potential function selection and
handles the problem as a stochastic filtering problem. The problem
is mapped from $\mathbb{SE}\left(3\right)$ to vector form, using
the Rodriguez vector and the position vector, and then followed by
the definition of the pose problem in the sense of Stratonovich. The
proposed filter guarantees that the errors present in position and
Rodriguez vector estimates are semi-globally uniformly ultimately
bounded (SGUUB) in mean square, and that they converge to small neighborhood
of the origin in probability. Simulation results show the robustness
and effectiveness of the proposed filter in presence of high levels
of noise and bias associated with the velocity vector as well as body-frame
measurements.
\end{abstract}
\end{frontmatter}{}

\section{Introduction}

This paper concerns the problem of position and attitude estimation
of a rigid-body moving in 3D space which is commonly known as the
pose problem. Pose (attitude and position) estimation is a crucial
task in robotics and engineering applications. The attitude and position
can be reconstructed through a set of vector measurements with respect
to body and inertial frames of reference. In general, the main objective
of pose estimation problem is to minimize the cost function similar
to Wahba's problem \cite{wahba1965least}. The approach applied in
\cite{wahba1965least} was purely algebraic, whereas other algorithms
used singular value decomposition to obtain comparable static solution
\cite{markley1988attitude}. However, the set of vectorial measurements
is susceptible to uncertainties such as slowly time-variant bias and
random noise components. Therefore, the static solutions proposed
in \cite{wahba1965least,markley1988attitude} give poor results. Traditionally,
the attitude estimation problem has been handled using Gaussian filters
or nonlinear deterministic filters which aimed to converge any initialized
estimate to the solution \cite{hashim2018Stoch,crassidis2007survey,hashim2018Conf1}. The family of Gaussian
attitude filters includes Kalman filter (KF) \cite{choukroun2006novel},
extended KF (EKF) \cite{lefferts1982kalman,liu2016vehicle}, multiplicative
EKF (MEKF) \cite{markley2003attitude}, and invariant EKF (IEKF) \cite{barrau2017invariant}.
A good survey of Gaussian attitude filters can be found in \cite{crassidis2007survey}.
Gaussian attitude filters often consider the unit quaternion in attitude
representation and go through liberalizations. From the other side,
the attitude problem is naturally nonlinear and nonlinear deterministic
attitude filters can be developed directly on the Special Orthogonal
Group $\mathbb{SO}\left(3\right)$ as a deterministic problem
(for example \cite{mahony2008nonlinear,grip2012attitude,liu2018complementary}).
In fact, nonlinear deterministic attitude filters are simpler in derivation
and representation. In addition, they require less computational power
and demonstrate better tracking performance in comparison with Gaussian
filters \cite{crassidis2007survey,mahony2008nonlinear,hashim2018Stoch}.
Therefore, it is better to address the attitude and attitude-position
filtering in a nonlinear sense.

Inertial measurement units (IMUs) have a prominent role in enriching
the research of attitude estimation. These units are inexpensive fostering
the researchers to propose nonlinear deterministic filters on $\mathbb{SO}\left(3\right)$
\cite{mahony2008nonlinear,grip2012attitude,liu2018complementary}.
Attitude estimation is an essential part of the pose estimation problem
and a critical task in the estimation process. Accordingly, the pose
estimation problem can be modeled and solved using a nonlinear deterministic
filter evolved on the Special Euclidean Group $\mathbb{SE}\left(3\right)$.
Recently, the design of pose filters received considerable attention
\cite{rehbinder2003pose,baldwin2007complementary,baldwin2009nonlinear,hua2011observer,vasconcelos2010nonlinear,dominguez2017simultaneous}.
A computer vision system that employs a monocular camera with IMUs
was developed for pose estimation \cite{rehbinder2003pose,baldwin2007complementary}.
The filter in \cite{baldwin2007complementary} evolved directly on
$\mathbb{SE}\left(3\right)$ and has been proven to be exponentially
stable. However, the filter requires attitude and position reconstruction
for implementation. Later, the nonlinear complementary filter that
evolved directly on $\mathbb{SE}\left(3\right)$ in \cite{baldwin2007complementary}
was modified using vectorial measurements without the need of attitude
and position reconstruction \cite{baldwin2009nonlinear,hua2011observer}.
For a good overview of pose estimation on $\mathbb{SE}\left(3\right)$
the reader is advised to visit \cite{blanco2010tutorial}. Despite
the simplicity of the filter design in \cite{baldwin2007complementary,baldwin2009nonlinear,hua2011observer},
simulation results showed high sensitivity to noise and bias introduced
in the measurements. Moreover, pose estimators such as \cite{rehbinder2003pose,baldwin2007complementary,baldwin2009nonlinear,hua2011observer,vasconcelos2010nonlinear}
disregard the noise in the filter design assuming only constant bias
introduced in the measuring process. Therefore, successful spacecraft
control applications, such as, \cite{ye2017robust,xu2017fuzzy,mohamed2014improved}
cannot be achieved without pose filters robust against uncertain measurements.

Therefore, in order to develop successful pose estimator, we need
to realize that 
\begin{enumerate}
	\item[1)] the pose problem is naturally nonlinear on $\mathbb{SE}\left(3\right)$;
	and 
	\item[2)]  the true pose kinematics rely on angular and translational velocity.
\end{enumerate}
However, the velocity vector is subject to slowly time-variant bias
and random noise components. Hence, in this work a nonlinear stochastic
position and attitude filter is developed on $\mathbb{SE}\left(3\right)$
in the sense of Stratonovich \cite{stratonovich1967topics}. The problem
is mapped from $\mathbb{SE}\left(3\right)$ to vector form which includes
position and Rodriquez vector such that $X:\mathbb{SE}\left(3\right)\rightarrow\mathbb{R}^{6}$.
In the case where the velocity measurements are corrupted with noise,
the aim is 
\begin{enumerate}
	\item[1)] to steer the error vectors towards an arbitrarily small neighborhood
	of the origin in probability; 
	\item[2)] to attenuate the noise impact for known or unknown bounded covariance;
	and 
	\item[3)] to show that the error in $X$ and estimates is semi-globally uniformly
	ultimately bounded (SGUUB) in mean square.
\end{enumerate}
The rest of the paper is organized as follows: Section \ref{sec:SE3STCH_Math-Notations}
presents an overview of mathematical notation, mapping from $\mathbb{SO}\left(3\right)$
to angle-axis and Rodriguez vector parameterization, $\mathbb{SE}\left(3\right)$
properties and some helpful properties for the nonlinear stochastic
position and attitude filter design on $\mathbb{SE}\left(3\right)$.
Pose estimation dynamic problem in the stochastic sense is presented
in Section \ref{sec:SE3STCH_Problem-Formulation-in}. The nonlinear
stochastic filter on $\mathbb{SE}\left(3\right)$ and the stability
analysis are presented in Section \ref{sec:SE3STCH_Stochastic-Complementary-Filters}.
Section \ref{sec:SE3STCH_Simulations} demonstrates numerical results
and shows the output performance of the proposed stochastic filter.
Finally, Section \ref{sec:SE3STCH_Conclusion} draws a conclusion
of this work.

\section{Mathematical Notation and Background \label{sec:SE3STCH_Math-Notations}}

Throughout this paper, $\mathbb{R}_{+}$ denotes the set of nonnegative
real numbers. $\mathbb{R}^{n}$ is the real $n$-dimensional space,
$\mathbb{R}^{n\times m}$ denotes the real $n\times m$ dimensional
space. For $x\in\mathbb{R}^{n}$, the Euclidean norm is defined by
$\left\Vert x\right\Vert =\sqrt{x^{\top}x}$ where $^{\top}$ is the
transpose of the associated component. $\mathcal{C}^{n}$ denotes
the set of functions with continuous $n$th partial derivatives. $\mathcal{K}$
denotes a set of continuous and strictly increasing functions such
that $\gamma:\mathbb{R}_{+}\rightarrow\mathbb{R}_{+}$ and vanishes
only at zero. $\mathcal{K}_{\infty}$ denotes a set of continuous
and strictly increasing functions which belong to class $\mathcal{K}$
and is unbounded. $\mathbb{P}\left\{ \cdot\right\} $ is a probability
and $\mathbb{E}\left[\cdot\right]$ is an expected value of the associated
component. $\lambda\left(\cdot\right)$ is the set of singular values
of associated matrix with $\underline{\lambda}\left(\cdot\right)$
being the minimum value. Also, $\mathbf{I}_{n}$ denotes identity
matrix with $n$-by-$n$ dimensions, $\underline{\mathbf{0}}_{n}=\left[0,\ldots,0\right]^{\top}\in\mathbb{R}^{n}$
is a zero vector with $n$ rows and one column, and $\underline{\mathbf{1}}_{n}=\left[1,\ldots,1\right]^{\top}\in\mathbb{R}^{n}$.
$V$ is a potential function, and for $V\left(x\right)$ we have $V_{x}=\partial V/\partial x$
and $V_{xx}=\partial^{2}V/\partial x^{2}$.

Notation for frames is as follows: $\left\{ \mathcal{B}\right\} $
denotes the body-frame and $\left\{ \mathcal{I}\right\} $ denotes
the inertial-frame. Let $\mathbb{GL}\left(3\right)$ denote the 3
dimensional general linear group. $\mathbb{GL}\left(3\right)$ is
a Lie group characterized by smooth multiplication and inversion.
The set of orthogonal group, $\mathbb{O}\left(3\right)$ is a subgroup
of the general linear group and is defined by 
\[
\mathbb{O}\left(3\right)=\left\{ \left.M\in\mathbb{R}^{3\times3}\right|M^{\top}M=MM^{\top}=\mathbf{I}_{3}\right\} 
\]
where $\mathbf{I}_{3}$ is the identity matrix. $\mathbb{SO}\left(3\right)$
denotes the Special Orthogonal Group and is a subgroup of the orthogonal
group and the general linear group. The attitude of a rigid body
is denoted by a rotational matrix $R$, and is defined as follows
\[
\mathbb{SO}\left(3\right)=\left\{ \left.R\in\mathbb{R}^{3\times3}\right|RR^{\top}=R^{\top}R=\mathbf{I}_{3}\text{, }{\rm det}\left(R\right)=+1\right\} 
\]
where ${\rm det\left(\cdot\right)}$ is the determinant of the associated
matrix. Let $\mathbb{SE}\left(3\right)$ denote the Special Euclidean
Group with $\mathbb{SE}\left(3\right)=\mathbb{SO}\left(3\right)\times\mathbb{R}^{3}$.
$\mathbb{SE}\left(3\right)$ is a subset of the affine group $\mathbb{GA}\left(3\right)=\mathbb{GL}\left(3\right)\times\mathbb{R}^{3}$
such that 
\[
\mathbb{SE}\left(3\right)=\left\{ \left.\boldsymbol{T}\in\mathbb{R}^{4\times4}\right|R\in\mathbb{SO}\left(3\right),P\in\mathbb{R}^{3}\right\} 
\]
where $\boldsymbol{T}\in\mathbb{SE}\left(3\right)$ is known as the
homogeneous representation or the transformation matrix of the rigid
body and is defined by 
\begin{equation}
	\boldsymbol{T}=\left[\begin{array}{cc}
		R & P\\
		\underline{\mathbf{0}}_{3}^{\top} & 1
	\end{array}\right]\in\mathbb{SE}\left(3\right)\label{eq:SE3STCH_T_matrix}
\end{equation}
with $P\in\mathbb{R}^{3}$ denoting position, $R\in\mathbb{SO}\left(3\right)$
denoting the attitude of the rigid-body in the space, and $\underline{\mathbf{0}}_{3}^{\top}$
being a zero row. The associated Lie-algebra of $\mathbb{SO}\left(3\right)$
is termed $\mathfrak{so}\left(3\right)$ and is defined by 
\[
\mathfrak{so}\left(3\right)=\left\{ \left.A\in\mathbb{R}^{3\times3}\right|A^{\top}=-A\right\} 
\]
with $A$ being the space of skew-symmetric matrices. Let us define
the map $\left[\cdot\right]_{\times}:\mathbb{R}^{3}\rightarrow\mathfrak{so}\left(3\right)$
such that 
\[
A=\left[\alpha\right]_{\times}=\left[\begin{array}{ccc}
0 & -\alpha_{3} & \alpha_{2}\\
\alpha_{3} & 0 & -\alpha_{1}\\
-\alpha_{2} & \alpha_{1} & 0
\end{array}\right]\in\mathfrak{so}\left(3\right),\hspace{1em}\alpha=\left[\begin{array}{c}
\alpha_{1}\\
\alpha_{2}\\
\alpha_{3}
\end{array}\right]
\]
For all $\alpha,\beta\in\mathbb{R}^{3}$, we have $\left[\alpha\right]_{\times}\beta=\alpha\times\beta$
where $\times$ is the cross product between the two vectors. Let
$\wedge$ be the wedge operator, and the wedge map $\left[\cdot\right]_{\wedge}:\mathbb{R}^{6}\rightarrow\mathfrak{se}\left(3\right)$
such that 
\[
\left[\mathcal{Y}\right]_{\wedge}=\left[\begin{array}{cc}
\left[y_{1}\right]_{\times} & y_{2}\\
\underline{\mathbf{0}}_{3}^{\top} & 0
\end{array}\right]\in\mathfrak{se}\left(3\right)
\]
where $\mathcal{Y}=\left[y_{1}^{\top},y_{2}^{\top}\right]^{\top}$
for $y_{1},y_{2}\in\mathbb{R}^{3}$. The Lie algebra of $\mathbb{SE}\left(3\right)$
is denoted by $\mathfrak{se}\left(3\right)$ and given by 
\begin{align*}
	\mathfrak{se}\left(3\right) & =\left\{ \left.\left[\mathcal{Y}\right]_{\wedge}\in\mathbb{R}^{4\times4}\right|\exists y_{1},y_{2}\in\mathbb{R}^{3}:\left[\mathcal{Y}\right]_{\wedge}=\left[\begin{array}{cc}
		\left[y_{1}\right]_{\times} & y_{2}\\
		\underline{\mathbf{0}}_{3}^{\top} & 0
	\end{array}\right]\right\} 
\end{align*}
Let the $\mathbf{vex}$ operator be the inverse of $\left[\cdot\right]_{\times}$,
denoted by $\mathbf{vex}:\mathfrak{so}\left(3\right)\rightarrow\mathbb{R}^{3}$
such that for $\alpha\in\mathbb{R}^{3}$ and $A=\left[\alpha\right]_{\times}\in\mathfrak{so}\left(3\right)$
we have 
\[
\mathbf{vex}\left(A\right)=\mathbf{vex}\left(\left[\alpha\right]_{\times}\right)=\alpha\in\mathbb{R}^{3}
\]
Let $\boldsymbol{\mathcal{P}}_{a}$ denote the anti-symmetric projection
operator on the Lie-algebra $\mathfrak{so}\left(3\right)$, defined
by $\boldsymbol{\mathcal{P}}_{a}:\mathbb{R}^{3\times3}\rightarrow\mathfrak{so}\left(3\right)$
such that 
\begin{equation}
	\boldsymbol{\mathcal{P}}_{a}\left(M\right)=\frac{1}{2}\left(M-M^{\top}\right)\in\mathfrak{so}\left(3\right),\,M\in\mathbb{R}^{3\times3}\label{eq:SE3STCH_Pa}
\end{equation}
Let us define $\boldsymbol{\Upsilon}_{a}\left(\cdot\right)$ as the
composition mapping such that $\boldsymbol{\Upsilon}_{a}=\mathbf{vex}\circ\boldsymbol{\mathcal{P}}_{a}$.
Hence, $\boldsymbol{\Upsilon}_{a}\left(M\right)$ can be expressed
for $M\in\mathbb{R}^{3\times3}$ as 
\begin{equation}
	\boldsymbol{\Upsilon}_{a}\left(M\right)=\mathbf{vex}\left(\boldsymbol{\mathcal{P}}_{a}\left(M\right)\right)\in\mathbb{R}^{3}\label{eq:SE3STCH_VEX_a}
\end{equation}
Consider $\boldsymbol{\mathcal{P}}:\mathbb{R}^{4\times4}\rightarrow\mathfrak{se}\left(3\right)$
denoting the projection operator on the space of the Lie algebra $\mathfrak{se}\left(3\right)$
such that for $\mathcal{M}=\left[\begin{array}{cc}
M & m_{x}\\
m_{y}^{\top} & m_{z}
\end{array}\right]\in\mathbb{R}^{4\times4}$ with $M\in\mathbb{R}^{3\times3}$, $m_{x},m_{y}\in\mathbb{R}^{3}$
and $m_{z}\in\mathbb{R}$, we have 
\begin{equation}
	\boldsymbol{\mathcal{P}}\left(\mathcal{M}\right)=\boldsymbol{\mathcal{P}}\left(\left[\begin{array}{cc}
		M & m_{x}\\
		m_{y}^{\top} & m_{z}
	\end{array}\right]\right)=\left[\begin{array}{cc}
		\boldsymbol{\mathcal{P}}_{a}\left(M\right) & m_{x}\\
		\underline{\mathbf{0}}_{3}^{\top} & 0
	\end{array}\right]\in\mathfrak{se}\left(3\right)\label{eq:SE3STCH_P}
\end{equation}
For any $\mathcal{M}\in\mathbb{R}^{4\times4}$, we define the operator
$\boldsymbol{\Upsilon}\left(\cdot\right)$ as follows 
\begin{equation}
	\boldsymbol{\Upsilon}\left(\mathcal{M}\right)=\left[\begin{array}{c}
		\boldsymbol{\Upsilon}_{a}\left(M\right)\\
		m_{x}
	\end{array}\right]\in\mathbb{R}^{6}\label{eq:SE3STCH_Vex}
\end{equation}
The normalized Euclidean distance of a rotation matrix on $\mathbb{SO}\left(3\right)$
is defined by 
\begin{equation}
	\left\Vert R\right\Vert _{I}=\frac{1}{4}{\rm Tr}\left\{ \mathbf{I}_{3}-R\right\} \label{eq:SE3STCH_Ecul_Dist}
\end{equation}
such that ${\rm Tr}\left\{ \cdot\right\} $ is the trace of the associated
matrix, while the normalized Euclidean distance of $R\in\mathbb{SO}\left(3\right)$
is $\left\Vert R\right\Vert _{I}\in\left[0,1\right]$. The orientation
of a rigid-body rotating in a 3D space can be established according
to its angle of rotation $\alpha\in\mathbb{R}$ and its axis parameterization
$u\in\mathbb{R}^{3}$ \cite{shuster1993survey}. Such parameterization
is termed angle-axis parameterization. Mapping from angle-axis parameterization
to $\mathbb{SO}\left(3\right)$ is given by $\mathcal{R}_{\alpha}:\mathbb{R}\times\mathbb{R}^{3}\rightarrow\mathbb{SO}\left(3\right)$
such that 
\begin{equation}
	\mathcal{R}_{\alpha}\left(\alpha,u\right)=\mathbf{I}_{3}+\sin\left(\alpha\right)\left[u\right]_{\times}+\left(1-\cos\left(\alpha\right)\right)\left[u\right]_{\times}^{2}\in\mathbb{SO}\left(3\right)\label{eq:SE3STCH_att_ang}
\end{equation}
In the same spirit, the orientation of a rigid-body can be constructed
by Rodriguez parameters vector. Mapping from Rodriguez vector parameterization
to $\mathbb{SO}\left(3\right)$ is defined by $\mathcal{R}_{\rho}:\mathbb{R}^{3}\rightarrow\mathbb{SO}\left(3\right)$
such that

\begin{align}
	\mathcal{R}_{\rho}\left(\rho\right)= & \frac{1}{1+\left\Vert \rho\right\Vert ^{2}}\left(\left(1-\left\Vert \rho\right\Vert ^{2}\right)\mathbf{I}_{3}+2\rho\rho^{\top}+2\left[\rho\right]_{\times}\right)\in\mathbb{SO}\left(3\right)\label{eq:SE3STCH_SO3_Rodr}
\end{align}
One can obtain the normalized Euclidean distance in \eqref{eq:SE3STCH_Ecul_Dist}
as a function of Rodriguez parameters vector substituting \eqref{eq:SE3STCH_SO3_Rodr}
into \eqref{eq:SE3STCH_Ecul_Dist} to yield 
\begin{equation}
	\left\Vert R\right\Vert _{I}=\frac{1}{4}{\rm Tr}\left\{ \mathbf{I}_{3}-R\right\} =\frac{\left\Vert \rho\right\Vert ^{2}}{1+\left\Vert \rho\right\Vert ^{2}}\in\left[0,1\right]\label{eq:SE3STCH_TR2}
\end{equation}
Also, the anti-symmetric projection operator of the attitude $R$,
denoted by $\boldsymbol{\mathcal{P}}_{a}\left(R\right)$, can be defined
in terms of Rodriguez parameters vector from \eqref{eq:SE3STCH_SO3_Rodr}
as 
\begin{align*}
	\boldsymbol{\mathcal{P}}_{a}\left(R\right)= & 2\frac{1}{1+\left\Vert \rho\right\Vert ^{2}}\left[\rho\right]_{\times}\in\mathfrak{so}\left(3\right)
\end{align*}
Accordingly, the composition mapping $\boldsymbol{\Upsilon}_{a}\left(\cdot\right)$
of $\boldsymbol{\mathcal{P}}_{a}\left(R\right)$ in \eqref{eq:SE3STCH_Pa}
and \eqref{eq:SE3STCH_VEX_a} can be defined in terms of Rodriguez
parameters vector as 
\begin{equation}
	\boldsymbol{\Upsilon}_{a}\left(R\right)=\mathbf{vex}\left(\boldsymbol{\mathcal{P}}_{a}\left(R\right)\right)=2\frac{\rho}{1+\left\Vert \rho\right\Vert ^{2}}\in\mathbb{R}^{3}\label{eq:SE3STCH_VEX_Pa}
\end{equation}
from \eqref{eq:SE3STCH_TR2} and \eqref{eq:SE3STCH_VEX_Pa}, it follows
\begin{align}
	\left\Vert \boldsymbol{\Upsilon}_{a}\left(R\right)\right\Vert ^{2} & =4\frac{\left\Vert \rho\right\Vert ^{2}}{\left(1+\left\Vert \rho\right\Vert ^{2}\right)^{2}}\nonumber \\
	& =4\left(1-\left\Vert R\right\Vert _{I}\right)\left\Vert R\right\Vert _{I}\label{eq:SE3STCH_VEX2_Pa}
\end{align}
Let us consider the transformation matrix in \eqref{eq:SE3STCH_T_matrix}
with $\boldsymbol{T}\in\mathbb{SE}\left(3\right)$. The adjoint map
for any $\boldsymbol{T}\in\mathbb{SE}\left(3\right)$ and $\mathcal{M}\in\mathfrak{se}\left(3\right)$
is given by 
\begin{equation}
	\mathbf{Ad}_{\boldsymbol{T}}\left(\mathcal{M}\right)=\boldsymbol{T}\mathcal{M}\boldsymbol{T}^{-1}\in\mathfrak{se}\left(3\right)\label{eq:SE3STCH_Adj1}
\end{equation}
Let us define another adjoint map for any $\boldsymbol{T}\in\mathbb{SE}\left(3\right)$
by 
\begin{equation}
	\breve{\overline{\mathbf{Ad}}}_{\boldsymbol{T}}=\left[\begin{array}{cc}
		R & \mathbf{0}_{3\times3}\\
		\left[P\right]_{\times}R & R
	\end{array}\right]\in\mathbb{R}^{6\times6}\label{eq:SE3STCH_Adj2}
\end{equation}
One can easily verify that the vex operator in \eqref{eq:SE3STCH_Vex}
can be combined with the results in \eqref{eq:SE3STCH_Adj1} and \eqref{eq:SE3STCH_Adj2}
to show 
\[
\mathbf{\Upsilon}\left(\mathbf{Ad}_{\boldsymbol{T}}\left(\mathcal{M}\right)\right)=\breve{\overline{\mathbf{Ad}}}_{\boldsymbol{T}}\mathbf{\Upsilon}\left(\mathcal{M}\right)\in\mathbb{R}^{6}
\]
thus 
\begin{equation}
	\boldsymbol{T}\left[\mathcal{Y}\right]_{\wedge}\boldsymbol{T}^{-1}=\left[\breve{\overline{\mathbf{Ad}}}_{\boldsymbol{T}}\mathcal{Y}\right]_{\wedge}\in\mathbb{SE}\left(3\right),\hspace{1em}\mathcal{Y}\in{\rm \mathbb{R}}^{6},\boldsymbol{T}\in\mathbb{SE}\left(3\right)\label{eq:SE3STCH_Identity5}
\end{equation}
which will be useful for the filter derivation and further analysis.
Finally, the following identities will be used in the subsequent derivations
\begin{align}
	-\left[\beta\right]_{\times}\left[\alpha\right]_{\times} & =\left(\beta^{\top}\alpha\right)\mathbf{I}_{3}-\alpha\beta^{\top},\quad\alpha,\beta\in{\rm \mathbb{R}}^{3}\label{eq:SE3STCH_Identity1}\\
	\left[R\alpha\right]_{\times} & =R\left[\alpha\right]_{\times}R^{\top},\quad R\in\mathbb{SO}\left(3\right),\alpha\in\mathbb{R}^{3}\label{eq:SE3STCH_Identity2}\\
	\breve{\overline{\mathbf{Ad}}}_{\boldsymbol{T}_{1}\boldsymbol{T}_{2}} & =\breve{\overline{\mathbf{Ad}}}_{\boldsymbol{T}_{1}}\breve{\overline{\mathbf{Ad}}}_{\boldsymbol{T}_{2}},\hspace{1em}\boldsymbol{T}_{1},\boldsymbol{T}_{2}\in\mathbb{SE}\left(3\right)\label{eq:SE3STCH_Identity3}\\
	\breve{\overline{\mathbf{Ad}}}_{\boldsymbol{T}}\breve{\overline{\mathbf{Ad}}}_{\boldsymbol{T}^{-1}} & =\breve{\overline{\mathbf{Ad}}}_{\boldsymbol{T}^{-1}}\breve{\overline{\mathbf{Ad}}}_{\boldsymbol{T}}=\mathbf{I}_{6},\hspace{1em}\boldsymbol{T}\in\mathbb{SE}\left(3\right)\label{eq:SE3STCH_Identity4}
\end{align}

\section{Problem Formulation in Stochastic Sense \label{sec:SE3STCH_Problem-Formulation-in}}

The orientation of a rigid-body rotating in 3D space $R\in\mathbb{SO}\left(3\right)$
is normally defined in terms of the body-frame $R\in\left\{ \mathcal{B}\right\} $
relative to the inertial-frame $\left\{ \mathcal{I}\right\} $. Let
$P\in\mathbb{R}^{3}$ be the position of the rigid-body measured on
the inertial-frame $P\in\left\{ \mathcal{I}\right\} $. Thereby, this
work concerns position as well as attitude estimation of a rigid-body
moving and rotating in 3D space. Consider the homogeneous transformation
matrix given by 
\begin{equation}
	\boldsymbol{T}=\left[\begin{array}{cc}
		R & P\\
		\underline{\mathbf{0}}_{3}^{\top} & 1
	\end{array}\right]\in\mathbb{SE}\left(3\right)\label{eq:SE3STCH_T_matrix2}
\end{equation}
Let $\Omega\in\mathbb{R}^{3}$ and $V\in\mathbb{R}^{3}$ be angular
and translational velocity of a moving rigid-body attached to the
body-frame, respectively, for all $\Omega,V\in\left\{ \mathcal{B}\right\} $.
Hence, the dynamics of the homogeneous transformation matrix $\boldsymbol{T}$
are expressed by 
\begin{align}
	\dot{P} & =RV\nonumber \\
	\dot{R} & =R\left[\Omega\right]_{\times}\label{eq:SE3STCH_R_dyn}\\
	\dot{\boldsymbol{T}} & =\boldsymbol{T}\left[\mathcal{Y}\right]_{\wedge}\label{eq:SE3STCH_T_Dynamics}
\end{align}
where $\mathcal{Y}=\left[\Omega^{\top},V^{\top}\right]^{\top}\in\mathbb{R}^{6}$
is the group velocity vector expressed relative to the body-frame.
The homogeneous transformation matrix $\boldsymbol{T}$ can be reconstructed
through a set of known vectors in the inertial-frame and their measurements
in the body-frame. Let the superscript $\mathcal{B}$ and $\mathcal{I}$
denote the associated body-frame and inertial-frame of the component,
respectively. The pose estimation problem is illustrated in Figure
\ref{fig:SE3STCH_1}. 
\begin{center}
	\begin{figure}[h]
		\centering{}\includegraphics[scale=0.6]{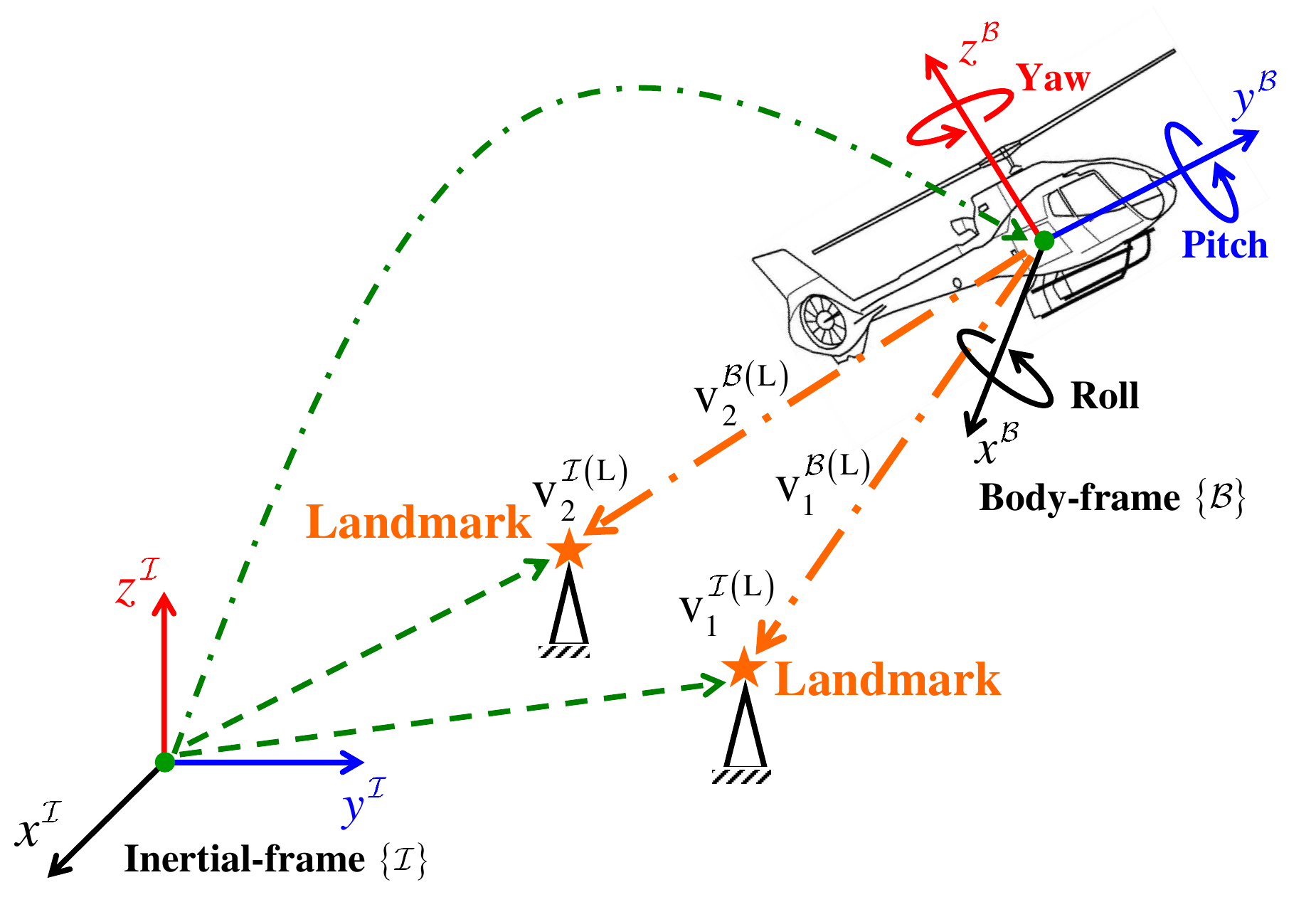}\caption{Pose estimation problem of a rigid-body moving in 3D space.}
		\label{fig:SE3STCH_1} 
	\end{figure}
	\par\end{center}

Assume that there exists a number of feature points or landmarks denoted
by $N_{{\rm L}}$ such that 
\begin{equation}
	{\rm v}_{i}^{\mathcal{B}\left({\rm L}\right)}=R^{\top}\left({\rm v}_{i}^{\mathcal{I}\left({\rm L}\right)}-P\right)+{\rm b}_{i}^{\mathcal{B}\left({\rm L}\right)}+\omega_{i}^{\mathcal{B}\left({\rm L}\right)}\label{eq:SE3STCH_Vec_Landmark}
\end{equation}
with ${\rm v}_{i}^{\mathcal{B}\left({\rm L}\right)}\in\mathbb{R}^{3}$
being the landmark measurement in the body-frame and ${\rm v}_{i}^{\mathcal{I}\left({\rm L}\right)}\in\mathbb{R}^{3}$
being a known constant feature in the inertial-frame for all $i=1,\ldots,N_{{\rm L}}$.
Also, ${\rm b}_{i}^{\mathcal{B}\left({\rm L}\right)}\in\mathbb{R}^{3}$
and $\omega_{i}^{\mathcal{B}\left({\rm L}\right)}\in\mathbb{R}^{3}$
are unknown bias and noise vectors attached to the $i$th measurement
for all $i=1,\ldots,N_{{\rm L}}$. The position $P$ can be simply
constructed if the attitude matrix $R$ is available. Let us denote
the set of vectors associated with landmarks by 
\begin{align}
	{\rm v}^{\mathcal{B}\left({\rm L}\right)} & =\left[{\rm v}_{1}^{\mathcal{B}\left({\rm L}\right)},\ldots,{\rm v}_{N_{{\rm L}}}^{\mathcal{B}\left({\rm L}\right)}\right]\in\left\{ \mathcal{B}\right\} \nonumber \\
	{\rm v}^{\mathcal{I}\left({\rm L}\right)} & =\left[{\rm v}_{1}^{\mathcal{I}\left({\rm L}\right)},\ldots,{\rm v}_{N_{{\rm L}}}^{\mathcal{I}\left({\rm L}\right)}\right]\in\left\{ \mathcal{I}\right\} \label{eq:SE3STCH_Set_L}
\end{align}
A weighted geometric center is considered for the case of more than
one landmark is available for measurement. The center is given by
\begin{align}
	P_{c}^{\mathcal{I}} & =\frac{1}{\sum_{i=1}^{N_{{\rm L}}}k_{i}^{{\rm L}}}\sum_{i=1}^{N_{{\rm L}}}k_{i}^{{\rm L}}{\rm v}_{i}^{\mathcal{I}\left({\rm L}\right)}\label{eq:SE3STCH_Center_Landmark_I}\\
	P_{c}^{\mathcal{B}} & =\frac{1}{\sum_{i=1}^{N_{{\rm L}}}k_{i}^{{\rm L}}}\sum_{i=1}^{N_{{\rm L}}}k_{i}^{{\rm L}}{\rm v}_{i}^{\mathcal{B}\left({\rm L}\right)}\label{eq:SE3STCH_Center_Landmark_B}
\end{align}
with $k_{i}^{{\rm L}}$ refers to the confidence level of the $i$th
measurement. On the other side, the attitude matrix $R$ can be obtained
through a set of $N_{{\rm R}}$-known non-collinear vectors. The $N_{{\rm R}}$
vectors are measured in the moving frame $\left\{ \mathcal{B}\right\} $.
Let ${\rm v}_{i}^{\mathcal{B}\left({\rm R}\right)}\in\mathbb{R}^{3}$
be a measured vector in the body-frame such that the $i$th body-frame
vector is given by 
\begin{equation}
	{\rm v}_{i}^{\mathcal{B}\left({\rm R}\right)}=R^{\top}{\rm v}_{i}^{\mathcal{I}\left({\rm R}\right)}+{\rm b}_{i}^{\mathcal{B}\left({\rm R}\right)}+\omega_{i}^{\mathcal{B}\left({\rm R}\right)}\label{eq:SE3STCH_Vect_R}
\end{equation}
where ${\rm v}_{i}^{\mathcal{I}\left({\rm R}\right)}$ refers to the
known vector $i$ in the inertial-frame for $i=1,2,\ldots,N_{{\rm R}}$.
${\rm b}_{i}^{\mathcal{B}\left({\rm R}\right)}$ and $\omega_{i}^{\mathcal{B}\left({\rm R}\right)}$
represent the unknown bias and noise components attached to the $i$th
measurement, respectively, for all ${\rm b}_{i}^{\mathcal{B}\left({\rm R}\right)},\omega_{i}^{\mathcal{B}\left({\rm R}\right)}\in\mathbb{R}^{3}$.
Let us denote the set of vectors associated with attitude reconstruction
by 
\begin{align}
	{\rm v}^{\mathcal{B}\left({\rm R}\right)} & =\left[{\rm v}_{1}^{\mathcal{B}\left({\rm R}\right)},\ldots,{\rm v}_{N_{{\rm R}}}^{\mathcal{B}\left({\rm R}\right)}\right]\in\left\{ \mathcal{B}\right\} \nonumber \\
	{\rm v}^{\mathcal{I}\left({\rm R}\right)} & =\left[{\rm v}_{1}^{\mathcal{I}\left({\rm R}\right)},\ldots,{\rm v}_{N_{{\rm R}}}^{\mathcal{I}\left({\rm R}\right)}\right]\in\left\{ \mathcal{I}\right\} \label{eq:SE3STCH_Set_R}
\end{align}
\begin{assum}
	\label{Assum:SE3STCH_1} At least one feature point is available for
	measurements \eqref{eq:SE3STCH_Vec_Landmark} with $N_{{\rm L}}\geq1$,
	and three non-collinear vectors are available for measurements \eqref{eq:SE3STCH_Vect_R}
	with $N_{{\rm R}}\geq2$. In case when $N_{{\rm R}}=2$, the third
	vector can be obtained by ${\rm v}_{3}^{\mathcal{I}\left({\rm R}\right)}={\rm v}_{1}^{\mathcal{I}\left({\rm R}\right)}\times{\rm v}_{2}^{\mathcal{I}\left({\rm R}\right)}$
	and ${\rm v}_{3}^{\mathcal{B}\left({\rm R}\right)}={\rm v}_{1}^{\mathcal{B}\left({\rm R}\right)}\times{\rm v}_{2}^{\mathcal{B}\left({\rm R}\right)}$. 
\end{assum}
According to Assumption \ref{Assum:SE3STCH_1}, $N_{{\rm R}}\geq2$
means that the set of vectorial measurements in \eqref{eq:SE3STCH_Set_R}
is sufficient to have rank 3. The homogeneous transformation matrix
$\boldsymbol{T}$ can be reconstructed if Assumption \ref{Assum:SE3STCH_1}
is satisfied. It is common to obtain the normalized values of inertial
and body-frame measurements in \eqref{eq:SE3STCH_Vect_R} such that
\begin{equation}
	\upsilon_{i}^{\mathcal{I}\left({\rm R}\right)}=\frac{{\rm v}_{i}^{\mathcal{I}\left({\rm R}\right)}}{\left\Vert {\rm v}_{i}^{\mathcal{I}\left({\rm R}\right)}\right\Vert },\hspace{1em}\upsilon_{i}^{\mathcal{B}\left({\rm R}\right)}=\frac{{\rm v}_{i}^{\mathcal{B}\left({\rm R}\right)}}{\left\Vert {\rm v}_{i}^{\mathcal{B}\left({\rm R}\right)}\right\Vert }\label{eq:SE3STCH_Vector_norm}
\end{equation}
and the normalized set of \eqref{eq:SE3STCH_Vector_norm} is 
\begin{align}
	\upsilon^{\mathcal{B}\left({\rm R}\right)} & =\left[\upsilon_{1}^{\mathcal{B}\left({\rm R}\right)},\ldots,\upsilon_{N_{{\rm R}}}^{\mathcal{B}\left({\rm R}\right)}\right]\in\left\{ \mathcal{B}\right\} \nonumber \\
	\upsilon^{\mathcal{I}\left({\rm R}\right)} & =\left[\upsilon_{1}^{\mathcal{I}\left({\rm R}\right)},\ldots,\upsilon_{N_{{\rm R}}}^{\mathcal{I}\left({\rm R}\right)}\right]\in\left\{ \mathcal{I}\right\} \label{eq:SE3STCH_Set_R_Norm}
\end{align}
In that case, the attitude can be extracted knowing $\upsilon_{i}^{\mathcal{I}\left({\rm R}\right)}$
and $\upsilon_{i}^{\mathcal{B}\left({\rm R}\right)}$ instead of ${\rm v}_{i}^{\mathcal{I}\left({\rm R}\right)}$
and ${\rm v}_{i}^{\mathcal{B}\left({\rm R}\right)}$. Gyroscope obtains
the measurements of angular velocity in the body-frame $\left\{ \mathcal{B}\right\} $
and the measurement vector is defined by 
\begin{equation}
	\Omega_{m}=\Omega+b_{\Omega}+\omega_{\Omega}\in\left\{ \mathcal{B}\right\} \label{eq:SE3STCH_Angular}
\end{equation}
with $\Omega$ denoting the true value of angular velocity, $b_{\Omega}\in\mathbb{R}^{3}$
denoting the bias component which is unknown constant or slowly time-varying
vector, and $\omega_{\Omega}\in\mathbb{R}^{3}$ being the unknown
noise component attached to angular velocity measurements. Also, the
translational velocity is expressed in the body-frame and its measurement
is defined by 
\begin{equation}
	V_{m}=V+b_{V}+\omega_{V}\in\left\{ \mathcal{B}\right\} \label{eq:SE3STCH_V_Trans}
\end{equation}
where $V$ denotes the true value of the translational velocity, $b_{V}\in\mathbb{R}^{3}$
denotes the unknown bias component, and $\omega_{V}\in\mathbb{R}^{3}$
is the unknown noise component attached to translational velocity
measurements. Let the group of velocity measurements, bias and noise
vectors be defined by $\mathcal{Y}_{m}=\left[\Omega_{m}^{\top},V_{m}^{\top}\right]^{\top}$,
$b=\left[b_{\Omega}^{\top},b_{V}^{\top}\right]^{\top}$ and $\omega=\left[\omega_{\Omega}^{\top},\omega_{V}^{\top}\right]^{\top}$,
respectively, for all $\mathcal{Y}_{m},b,\omega\in\mathbb{R}^{6}$.
The noise vector $\omega$ is assumed to be Gaussian with zero mean.
The dynamics of \eqref{eq:SE3STCH_R_dyn} can be mapped to Rodriguez
vector and expressed as follows \cite{shuster1993survey} 
\begin{equation}
	\dot{\rho}=\frac{1}{2}\left(\mathbf{I}_{3}+\left[\rho\right]_{\times}+\rho\rho^{\top}\right)\Omega\label{eq:SE3STCH_Rod_dynam}
\end{equation}
Therefore, the dynamics of the homogeneous transformation matrix in
\eqref{eq:SE3STCH_T_Dynamics} can be mapped to vector form in the
sense of Rodriguez parameters from \eqref{eq:SE3STCH_Rod_dynam} and
\eqref{eq:SE3STCH_SO3_Rodr} as 
\begin{equation}
	\left[\begin{array}{c}
		\dot{\rho}\\
		\dot{P}
	\end{array}\right]=\left[\begin{array}{cc}
		\frac{\mathbf{I}_{3}+\left[\rho\right]_{\times}+\rho\rho^{\top}}{2} & \mathbf{0}_{3\times3}\\
		\mathbf{0}_{3\times3} & \mathcal{R}_{\rho}\left(\rho\right)
	\end{array}\right]\left[\begin{array}{c}
		\Omega\\
		V
	\end{array}\right]\label{eq:SE3STCH_Rod_Dynamics}
\end{equation}
where $\mathcal{R}_{\rho}\left(\rho\right)=R\in\mathbb{SO}\left(3\right)$
as given in \eqref{eq:SE3STCH_SO3_Rodr}. According to \eqref{eq:SE3STCH_Angular}
and \eqref{eq:SE3STCH_V_Trans}, the measurements of angular and translational
velocities are subject to noise and bias components. These components
are characterized by randomness and uncertainty. As such, random behavior and the randomness
in measurements could lead to unknown behavior \cite{hashim2018Stoch,hashim2017neuro,hashim2017adaptive}
and impair the whole estimation process. The dynamics of the homogeneous
transformation matrix in \eqref{eq:SE3STCH_T_Dynamics} become 
\begin{equation}
	\dot{\boldsymbol{T}}=\boldsymbol{T}\left[\begin{array}{c}
		\mathcal{Y}_{m}-b-\omega\end{array}\right]_{\wedge}\label{eq:SE3STCH_T_Dynam_Noise}
\end{equation}

\noindent In view of \eqref{eq:SE3STCH_T_Dynamics} and \eqref{eq:SE3STCH_Rod_Dynamics},
the dynamics in \eqref{eq:SE3STCH_T_Dynam_Noise} can be mapped in
the same sense and represented as 
\begin{equation}
	\left[\begin{array}{c}
		\dot{\rho}\\
		\dot{P}
	\end{array}\right]=\left[\begin{array}{cc}
		\frac{\mathbf{I}_{3}+\left[\rho\right]_{\times}+\rho\rho^{\top}}{2} & \mathbf{0}_{3\times3}\\
		\mathbf{0}_{3\times3} & \mathcal{R}_{\rho}\left(\rho\right)
	\end{array}\right]\left(\mathcal{Y}_{m}-b-\omega\right)\label{eq:SE3STCH_X_Noise}
\end{equation}
where $\omega$ is a continuous Gaussian random noise vector with
zero mean which is bounded. The derivative of any Gaussian process
yields a Gaussian process \cite{khasminskii1980stochastic,jazwinski2007stochastic}.
Hence, the vector $\omega$ can be written as a function of Brownian
motion process vector $d\beta/dt$ with $\beta\in\mathbb{R}^{6}$
such that 
\[
\omega=\mathcal{Q}\frac{d\beta}{dt}
\]
where $\beta=\left[\beta_{\Omega}^{\top},\beta_{V}^{\top}\right]^{\top}$
and $\mathcal{Q}\in\mathbb{R}^{6\times6}$ is a diagonal matrix whose
diagonal has unknown time-variant nonnegative components defined by
\[
\mathcal{Q}=\left[\begin{array}{cc}
\mathcal{Q}_{\Omega} & \mathbf{0}_{3\times3}\\
\mathbf{0}_{3\times3} & \mathcal{Q}_{V}
\end{array}\right]
\]
where $\mathcal{Q}_{\Omega}\in\mathbb{R}^{3\times3}$ is associated
with $\omega_{\Omega}$ and $\mathcal{Q}_{V}\in\mathbb{R}^{3\times3}$
is associated with $\omega_{V}$. In addition, $\mathcal{Q}^{2}=\mathcal{Q}\mathcal{Q}^{\top}$
is a covariance component associated with the noise vector $\omega$.
The properties of Brownian motion process are defined by \cite{jazwinski2007stochastic,ito1984lectures,deng2001stabilization}
\[
\mathbb{P}\left\{ \beta\left(0\right)=0\right\} =1,\hspace{1em}\mathbb{E}\left[d\beta/dt\right]=0,\hspace{1em}\mathbb{E}\left[\beta\right]=0
\]
Let the dynamics of the homogeneous transformation in \eqref{eq:SE3STCH_T_Dynamics}
be defined in the sense of Stratonovich \cite{stratonovich1967topics}
and substitute $\omega$ by $\mathcal{Q}d\beta/dt$. Accordingly,
the stochastic differential equation of \eqref{eq:SE3STCH_T_Dynamics}
can be expressed as 
\begin{equation}
	d\boldsymbol{T}=\boldsymbol{T}\left[\mathcal{Y}_{m}-b\right]_{\wedge}dt-\boldsymbol{T}\left[\mathcal{Q}d\beta\right]_{\wedge}\label{eq:SE3STCH_T_Ito}
\end{equation}

\noindent in view of \eqref{eq:SE3STCH_T_Dynam_Noise} and \eqref{eq:SE3STCH_X_Noise},
the stochastic differential equation in \eqref{eq:SE3STCH_T_Ito}
is given by 
\begin{align}
	\left[\begin{array}{c}
		d\rho\\
		dP
	\end{array}\right]= & \left[\begin{array}{cc}
		\frac{\mathbf{I}_{3}+\left[\rho\right]_{\times}+\rho\rho^{\top}}{2} & \mathbf{0}_{3\times3}\\
		\mathbf{0}_{3\times3} & \mathcal{R}_{\rho}\left(\rho\right)
	\end{array}\right]\left(\left(\mathcal{Y}_{m}-b\right)dt-\mathcal{Q}d\beta\right)\label{eq:SE3STCH_X_Ito}
\end{align}
Let us define 
\begin{align}
	dX & =f\left(\rho,b\right)dt-\mathcal{G}\left(\rho\right)\mathcal{Q}d\beta\label{eq:SE3STCH_X_Ito-1}\\
	\mathcal{G}\left(\rho\right) & =\left[\begin{array}{cc}
		g_{\rho} & \mathbf{0}_{3\times3}\\
		\mathbf{0}_{3\times3} & g_{P}
	\end{array}\right]=\left[\begin{array}{cc}
		\frac{\mathbf{I}_{3}+\left[\rho\right]_{\times}+\rho\rho^{\top}}{2} & \mathbf{0}_{3\times3}\\
		\mathbf{0}_{3\times3} & \mathcal{R}_{\rho}\left(\rho\right)
	\end{array}\right]\nonumber \\
	f\left(\rho,b\right) & =\mathcal{G}\left(\rho\right)\left(\mathcal{Y}_{m}-b\right)\nonumber 
\end{align}

\noindent with $X=\left[\rho^{\top},P^{\top}\right]^{\top}\in\mathbb{R}^{6}$,
$\mathcal{G}:\mathbb{R}^{3}\rightarrow\mathbb{R}^{6\times6}$ and
$f:\mathbb{R}^{3}\times\mathbb{R}^{6}\rightarrow\mathbb{R}^{6}$.
$\mathcal{G}\left(\rho\right)$ is locally Lipschitz in $\rho$ and
$f\left(\rho,b\right)$ is locally Lipschitz in $\rho$ and $b$.
Consequently, the dynamic system in \eqref{eq:SE3STCH_X_Ito} has
a solution on $t\in\left[t\left(0\right),T\right]\forall t\left(0\right)\leq T<\infty$
in the mean square sense and for any $\rho\left(t\right)$ and $P\left(t\right)$
such that $t\neq t\left(0\right)$, $X-X\left(0\right)$ is independent
of $\left\{ \beta\left(\tau\right),\tau\geq t\right\} ,\forall t\in\left[t\left(0\right),T\right]$
(Theorem 4.5 \cite{jazwinski2007stochastic}). The aim is to achieve
adaptive stabilization of an unknown constant bias and unknown time-variant
covariance matrix. Let $\sigma=\left[\sigma_{\Omega}^{\top},\sigma_{V}^{\top}\right]^{\top}\in\mathbb{R}^{6}$
with $\sigma_{\Omega},\sigma_{V}\in\mathbb{R}^{3}$ being the upper
bound of $\mathcal{Q}^{2}$ such that 
\begin{equation}
	\sigma=\left[{\rm max}\left\{ \mathcal{Q}_{\left(1,1\right)}^{2}\right\} ,{\rm max}\left\{ \mathcal{Q}_{\left(2,2\right)}^{2}\right\} ,\ldots,{\rm max}\left\{ \mathcal{Q}_{\left(6,6\right)}^{2}\right\} \right]^{\top}\label{eq:SE3STCH_g_factor}
\end{equation}
where ${\rm max}\left\{ \cdot\right\} $ is the maximum value of the
associated covariance element. 
\begin{assum}
	\label{Assum:SE3STCH_2} Both $b$ and $\sigma$ belong to a given
	compact set $\Delta$ and are upper bounded by a scalar $\Gamma$
	such that $\left\Vert \Delta\right\Vert \leq\Gamma<\infty$. 
\end{assum}
\begin{defn}
	\label{def:SE3STCH_1}\cite{ji2006adaptive} The trajectory $X=\left[\rho^{\top},P^{\top}\right]^{\top}$
	of the stochastic differential system in \eqref{eq:SE3STCH_X_Ito}
	is said to be semi-globally uniformly ultimately bounded (SGUUB) if
	for some compact set $\Xi\in\mathbb{R}^{6}$ and any $X\left(0\right)=X\left(t\left(0\right)\right)$,
	there exists a constant $\vartheta>0$, and a time constant $T=T\left(\vartheta,X\left(0\right)\right)$
	such that $\mathbb{E}\left[\left\Vert X\right\Vert \right]<\vartheta,\forall t>t\left(0\right)+T$. 
\end{defn}
\begin{defn}
	\label{def:SE3STCH_2}Consider the stochastic differential system
	in \eqref{eq:SE3STCH_X_Ito} with $X=\left[\rho^{\top},P^{\top}\right]^{\top}$.
	For a given function $V\left(X\right)\in\mathcal{C}^{2}$ the differential
	operator $\mathcal{L}V$ is given by 
	\[
	\mathcal{L}V\left(X\right)=V_{X}^{\top}f\left(\rho,b\right)+\frac{1}{2}{\rm Tr}\left\{ \mathcal{G}\left(\rho\right)\mathcal{Q}^{2}\mathcal{G}^{\top}\left(\rho\right)V_{XX}\right\} 
	\]
	such that $V_{X}=\partial V/\partial X$, and $V_{XX}=\partial^{2}V/\partial X^{2}$. 
\end{defn}
\begin{lem}
	\label{lem:SE3STCH_1} \cite{deng2001stabilization,ji2006adaptive,deng1997stochastic}
	Consider the dynamic system in \eqref{eq:SE3STCH_X_Ito} with potential
	function $V\in\mathcal{C}^{2}$, such that $V:\mathbb{R}^{6}\rightarrow\mathbb{R}_{+}$,
	class $\mathcal{K}_{\infty}$ function $\bar{\alpha}_{1}\left(\cdot\right)$
	and $\bar{\alpha}_{2}\left(\cdot\right)$, constants $c_{1}>0$ and
	$c_{2}\geq0$ and a nonnegative function $\mathbf{Z}\left(\left\Vert X\right\Vert \right)$
	such that 
	\begin{equation}
		\bar{\alpha}_{1}\left(\left\Vert X\right\Vert \right)\leq V\leq\bar{\alpha}_{2}\left(\left\Vert X\right\Vert \right)\label{eq:SE3STCH_Vfunction_Lyap}
	\end{equation}
	\begin{align}
		\mathcal{L}V\left(X\right)= & V_{X}^{\top}f\left(\rho,b\right)+\frac{1}{2}{\rm Tr}\left\{ \mathcal{G}\left(\rho\right)\mathcal{Q}^{2}\mathcal{G}^{\top}\left(\rho\right)V_{XX}\right\} \nonumber \\
		\leq & -c_{1}\mathbf{Z}\left(\left\Vert X\right\Vert \right)+c_{2}\label{eq:SE3STCH_dVfunction_Lyap}
	\end{align}
	then for $X\left(0\right)\in\mathbb{R}^{6}$, there exists almost
	a unique strong solution on $\left[0,\infty\right)$ for the dynamic
	system in \eqref{eq:SE3STCH_X_Ito}. The solution $X$ is bounded
	in probability such that 
	\begin{equation}
		\mathbb{E}\left[V\left(X\right)\right]\leq V\left(X\left(0\right)\right){\rm exp}\left(-c_{1}t\right)+\frac{c_{2}}{c_{1}}\label{eq:SE3STCH_EVfunction_Lyap}
	\end{equation}
	Moreover, if the inequality in \eqref{eq:SE3STCH_EVfunction_Lyap}
	holds, then $X$ in \eqref{eq:SE3STCH_X_Ito} is SGUUB in the mean
	square. Also, when $c_{2}=0$, $f\left(0,b\right)=0$, $\mathcal{G}\left(0\right)=0$,
	and $\mathbf{Z}\left(\left\Vert X\right\Vert \right)$ is continuous,
	the equilibrium point $X=0$ is globally asymptotically stable in
	probability and the solution of $X$ satisfies 
	\begin{equation}
		\mathbb{P}\left\{ \underset{t\rightarrow\infty}{{\rm lim}}\mathbf{Z}\left(\left\Vert X\right\Vert \right)=0\right\} =1,\hspace{1em}\forall X\left(0\right)\in\mathbb{R}^{6}\label{eq:SE3STCH_PdVfunction_Lyap}
	\end{equation}
\end{lem}
The proof of this lemma and the existence of a unique solution can
be found in \cite{deng2001stabilization}. For a rotation matrix $R\in\mathbb{SO}\left(3\right)$,
let us define $\mathcal{U}_{0}\subseteq\mathbb{SO}\left(3\right)\times\mathbb{R}^{3}$
by $\mathcal{U}_{0}=\left\{ \left.\left(R\left(0\right),P\left(0\right)\right)\right|{\rm Tr}\left\{ R\left(0\right)\right\} =-1,P\left(0\right)=\underline{\mathbf{0}}_{3}\right\} $.
The set $\mathcal{U}_{0}$ is forward invariant and unstable for the
dynamic system \eqref{eq:SE3STCH_R_dyn} and \eqref{eq:SE3STCH_T_Dynamics},
as ${\rm Tr}\left\{ R\left(0\right)\right\} =-1$ implies $\rho\left(0\right)=\infty$
\cite{shuster1993survey,hashim2018Stoch}. From almost any initial
condition such that $R\left(0\right)\notin\mathcal{U}_{0}$ or equivalently
$\rho\left(0\right)\in\mathbb{R}^{3}$, we have $-1<{\rm Tr}\left\{ R\left(0\right)\right\} \leq3$
and the trajectory of $X=\left[\rho^{\top},P^{\top}\right]^{\top}$
converges to the neighborhood of the equilibrium point conditioned
on the value of $c_{2}$ in \eqref{eq:SE3STCH_dVfunction_Lyap}. 
\begin{lem}
	\label{lem:SE3STCH_2} (Young's inequality) Let $x$ and $y$ be real
	values such that $x,y\in\mathbb{R}^{3}$. Then, for any positive real
	numbers $c$ and $d$ satisfying $\frac{1}{c}+\frac{1}{d}=1$ with
	appropriately small positive constant $\varepsilon$, the following
	inequality holds 
	\begin{equation}
		x^{\top}y\leq\left(1/c\right)\varepsilon^{c}\left\Vert x\right\Vert ^{c}+\left(1/d\right)\varepsilon^{-d}\left\Vert y\right\Vert ^{d}\label{eq:SE3STCH_lem_ineq}
	\end{equation}
\end{lem}

\section{Nonlinear Stochastic Complementary Filter on $\mathbb{SE}\left(3\right)$
	\label{sec:SE3STCH_Stochastic-Complementary-Filters}}

Let $\hat{\boldsymbol{T}}$ be the estimator of the homogeneous transformation
matrix $\boldsymbol{T}$ such that 
\[
\hat{\boldsymbol{T}}=\left[\begin{array}{cc}
\hat{R} & \hat{P}\\
\underline{\mathbf{0}}_{3}^{\top} & 1
\end{array}\right]\in\mathbb{SE}\left(3\right)
\]
The main purpose of this section is to design a pose estimator to
drive $\hat{\boldsymbol{T}}\rightarrow\boldsymbol{T}$. Let us define
the error in the estimation of the homogeneous transformation matrix
by 
\begin{align}
	\tilde{\boldsymbol{T}} & =\boldsymbol{T}\hat{\boldsymbol{T}}^{-1}=\left[\begin{array}{cc}
		R\hat{R}^{\top} & P-R\hat{R}^{\top}\hat{P}\\
		\underline{\mathbf{0}}_{3}^{\top} & 1
	\end{array}\right]=\left[\begin{array}{cc}
		\tilde{R} & \tilde{P}\\
		\underline{\mathbf{0}}_{3}^{\top} & 1
	\end{array}\right]\label{eq:SE3STCH_T_error}
\end{align}
with $\tilde{R}=R\hat{R}^{\top}$ and $\tilde{P}=P-\tilde{R}\hat{P}$.
Driving $\hat{\boldsymbol{T}}\rightarrow\boldsymbol{T}$ guarantees
that $\tilde{P}\rightarrow0$ and $\tilde{\rho}\rightarrow0$, where
$\tilde{P}$ is the position error associated with $\tilde{\boldsymbol{T}}$
and $\tilde{\rho}$ is the error of Rodriguez vector associated with
$\tilde{R}$ which is in turn associated with $\tilde{\boldsymbol{T}}$.
In this Section, a nonlinear deterministic filter on $\mathbb{SE}\left(3\right)$
is presented. This filter is subsequently modified into a nonlinear
stochastic filter evolved directly on $\mathbb{SE}\left(3\right)$.
The nonlinear stochastic filter is driven in the sense of Stratonovich.
For $\tilde{X}=\left[\tilde{\rho}^{\top},\tilde{P}^{\top}\right]^{\top}\in\mathbb{R}^{6}$,
the error vector $\tilde{X}$ is regulated to an arbitrarily small
neighborhood of the origin in the case where velocity vector measurements
$\mathcal{Y}_{m}$ are contaminated with constant bias and random
noise at each time instant. Let $\hat{b}$ and $\hat{\sigma}$ denote
estimates of unknown parameters $b$, and $\sigma$, respectively.
Let the error in vector $b$ and $\sigma$ be defined by 
\begin{align}
	\tilde{b} & =b-\hat{b}\label{eq:SE3STCH_b_tilde}\\
	\tilde{\sigma} & =\sigma-\hat{\sigma}\label{eq:SE3STCH_sigma_tilde_strat}
\end{align}

\subsection{Nonlinear Deterministic Pose Filter \label{subsec:SE3STCH_Nonlinear-Ito}}

The aim of this subsection is to study the behavior of nonlinear deterministic
pose filter evolved directly on $\mathbb{SE}\left(3\right)$ in presence
of noise in the velocity vector measurements $\mathcal{Y}_{m}$. The
attitude can be constructed algebraically given a set of measurements
in \eqref{eq:SE3STCH_Set_R} to form $R_{y}$, for example \cite{wahba1965least,markley1988attitude}.
However, $R_{y}$ is uncertain and significantly far from the true
$R$. The given set of measurements in \eqref{eq:SE3STCH_Set_R_Norm}
helps in finding $R_{y}$ and for a given landmark(s) we have $P_{y}=\frac{1}{\sum_{i=1}^{N_{{\rm L}}}k_{i}^{{\rm L}}}\sum_{i=1}^{N_{{\rm L}}}k_{i}^{{\rm L}}\left({\rm v}_{i}^{\mathcal{I}\left({\rm L}\right)}-R_{y}{\rm v}_{i}^{\mathcal{B}\left({\rm L}\right)}\right)$
and $\boldsymbol{T}_{y}=\left[\begin{array}{cc}
R_{y} & P_{y}\\
\underline{\mathbf{0}}_{3}^{\top} & 1
\end{array}\right]$. Hence, the filter design aims to use the given measured $\boldsymbol{T}_{y}$,
and the velocity measurements in \eqref{eq:SE3STCH_Angular}, and
\eqref{eq:SE3STCH_V_Trans} to obtain a good estimate of the true
$\boldsymbol{T}$. Consider the nonlinear deterministic pose filter
design 
\begin{eqnarray}
	\dot{\hat{\boldsymbol{T}}} & = & \hat{\boldsymbol{T}}\left[\mathcal{Y}_{m}-\hat{b}+k_{w}W\right]_{\wedge},\hspace{1em}\hat{\boldsymbol{T}}\left(0\right)\in\mathbb{SE}\left(3\right)\label{eq:SE3STCH_Test_Det}\\
	\dot{\hat{b}} & = & -\Gamma\breve{\overline{\mathbf{Ad}}}_{\hat{\boldsymbol{T}}}^{\top}\left[\begin{array}{cc}
		\left\Vert \tilde{R}\right\Vert _{I}\mathbf{I}_{3} & \mathbf{0}_{3\times3}\\
		\mathbf{0}_{3\times3} & 4\tilde{R}^{\top}
	\end{array}\right]\boldsymbol{\Upsilon}\left(\tilde{\boldsymbol{T}}\right)-k_{b}\Gamma\hat{b}\label{eq:SE3STCH_best_Det}\\
	W & = & k_{p}\breve{\overline{\mathbf{Ad}}}_{\hat{\boldsymbol{T}}}^{-1}\left[\begin{array}{cc}
		\frac{2-\left\Vert \tilde{R}\right\Vert _{I}}{1-\left\Vert \tilde{R}\right\Vert _{I}}\mathbf{I}_{3} & \mathbf{0}_{3\times3}\\
		\mathbf{0}_{3\times3} & \tilde{R}^{\top}
	\end{array}\right]\boldsymbol{\Upsilon}\left(\tilde{\boldsymbol{T}}\right)\label{eq:SE3STCH_Wcorr_Det}
\end{eqnarray}
where $\mathcal{Y}_{m}=\left[\Omega_{m}^{\top},V_{m}^{\top}\right]^{\top}$
is a measured vector of angular and translational velocity defined
in \eqref{eq:SE3STCH_Angular} and \eqref{eq:SE3STCH_V_Trans}, respectively,
with no noise attached to measurements $\left(\omega=0\right)$. $\hat{b}=\left[\hat{b}_{\Omega}^{\top},\hat{b}_{V}^{\top}\right]^{\top}\in\mathbb{R}^{6}$
is the estimate of the unknown bias vector $b$, $\tilde{\boldsymbol{T}}=\boldsymbol{T}_{y}\hat{\boldsymbol{T}}^{-1}$,
$\boldsymbol{\Upsilon}\left(\tilde{\boldsymbol{T}}\right)=\left[\boldsymbol{\Upsilon}_{a}^{\top}\left(\tilde{R}\right),\tilde{P}^{\top}\right]^{\top}$
as in \eqref{eq:SE3STCH_Vex}, $\boldsymbol{\Upsilon}_{a}\left(\tilde{R}\right)=\mathbf{vex}\left(\boldsymbol{\mathcal{P}}_{a}\left(\tilde{R}\right)\right)$,
and $\left\Vert \tilde{R}\right\Vert _{I}=\frac{1}{4}{\rm Tr}\left\{ \mathbf{I}_{3}-\tilde{R}\right\} $.
Also, $\breve{\overline{\mathbf{Ad}}}_{\hat{\boldsymbol{T}}}=\left[\begin{array}{cc}
\hat{R} & \mathbf{0}_{3\times3}\\
\left[\hat{P}\right]_{\times}\hat{R} & \hat{R}
\end{array}\right]$, $\Gamma=\left[\begin{array}{cc}
\Gamma_{\Omega} & \mathbf{0}_{3\times3}\\
\mathbf{0}_{3\times3} & \Gamma_{V}
\end{array}\right]=\gamma\mathbf{I}_{6}$, is an adaptation gain with $\Gamma_{\Omega},\Gamma_{V}\in\mathbb{R}^{3\times3}$,
$\gamma>0$, and $k_{b}$, $k_{p}$ and $k_{w}$ are positive constants. 
\begin{thm}
	\label{thm:SE3STCH_1}\textbf{ }Consider the homogeneous transformation
	matrix dynamics in \eqref{eq:SE3STCH_T_Dynamics} with velocity measurements
	$\mathcal{Y}_{m}$ in \eqref{eq:SE3STCH_Angular} and \eqref{eq:SE3STCH_V_Trans}.
	Let Assumption \ref{Assum:SE3STCH_1} hold and assume that the vector
	measurements in \eqref{eq:SE3STCH_Vect_R} are normalized to \eqref{eq:SE3STCH_Vector_norm}.
	Let $\boldsymbol{T}_{y}$ be reconstructed using the vector measurement
	in \eqref{eq:SE3STCH_Vec_Landmark} and \eqref{eq:SE3STCH_Vector_norm}
	, and be coupled with the observer in \eqref{eq:SE3STCH_Test_Det},
	\eqref{eq:SE3STCH_best_Det} and \eqref{eq:SE3STCH_Wcorr_Det}. In
	case when velocity vector measurements $\mathcal{Y}_{m}$ are subject
	to constant bias, no noise is introduced to the system $\left(\omega=0\right)$,
	$\tilde{X}\left(0\right)=\left[\tilde{\rho}\left(0\right)^{\top},\tilde{P}\left(0\right)^{\top}\right]^{\top}\in\mathbb{R}^{6}$,
	and $\tilde{X}\left(0\right)\neq\underline{\mathbf{0}}_{6}$, 1) the
	error vector $\tilde{X}$ is uniformly ultimately bounded for all
	$t\geq t\left(0\right)$; and 2) consequently $\left(\tilde{\boldsymbol{T}},\tilde{b}\right)$
	steers to the neighborhood of the equilibrium set $\mathcal{S}=\left\{ \left(\tilde{\boldsymbol{T}},\tilde{b}\right)\in\mathbb{SE}\left(3\right)\times\mathbb{R}^{6}:\tilde{\boldsymbol{T}}=\mathbf{I}_{4},\tilde{b}=\underline{\mathbf{0}}_{6}\right\} $. 
\end{thm}
\textbf{Proof.} Let the error in $b$ and $\tilde{\boldsymbol{T}}$
be defined as in \eqref{eq:SE3STCH_b_tilde}, and \eqref{eq:SE3STCH_T_error},
respectively. Therefore, the derivative of homogeneous transformation
matrix error in \eqref{eq:SE3STCH_T_error} can be expressed from
\eqref{eq:SE3STCH_T_Dynam_Noise} and \eqref{eq:SE3STCH_Test_Det}
as 
\begin{align}
	\dot{\tilde{\boldsymbol{T}}} & =\dot{\boldsymbol{T}}\hat{\boldsymbol{T}}^{-1}+\boldsymbol{T}\dot{\hat{\boldsymbol{T}}}^{-1}\nonumber \\
	& =\boldsymbol{T}\left[\mathcal{Y}_{m}-b\right]_{\wedge}\hat{\boldsymbol{T}}^{-1}-\boldsymbol{T}\left[\mathcal{Y}_{m}-\hat{b}+k_{w}W\right]_{\wedge}\hat{\boldsymbol{T}}^{-1}\nonumber \\
	& =\boldsymbol{T}\hat{\boldsymbol{T}}^{-1}\hat{\boldsymbol{T}}\left[-\tilde{b}-k_{w}W\right]_{\wedge}\hat{\boldsymbol{T}}^{-1}\nonumber \\
	& =-\tilde{\boldsymbol{T}}\left[\breve{\overline{\mathbf{Ad}}}_{\hat{\boldsymbol{T}}}\left(\tilde{b}+k_{w}W\right)\right]_{\wedge}\label{eq:SE3STCH_Terr_Det}
\end{align}
where $\dot{\hat{\boldsymbol{T}}}^{-1}=-\hat{\boldsymbol{T}}^{-1}\dot{\hat{\boldsymbol{T}}}\hat{\boldsymbol{T}}^{-1}$,
and $\tilde{b}=\left[\tilde{b}_{\Omega}^{\top},\tilde{b}_{V}^{\top}\right]^{\top}$.
Considering the math identity in \eqref{eq:SE3STCH_Identity5}, we
have $\hat{\boldsymbol{T}}\left[\tilde{b}\right]_{\wedge}\hat{\boldsymbol{T}}^{-1}=\left[\breve{\overline{\mathbf{Ad}}}_{\hat{\boldsymbol{T}}}\tilde{b}\right]_{\wedge}$.
For $\tilde{X}=\left[\tilde{\rho}^{\top},\tilde{P}^{\top}\right]^{\top}$,
and in view of the transformation of \eqref{eq:SE3STCH_T_Dynam_Noise}
into \eqref{eq:SE3STCH_X_Noise}, one may write \eqref{eq:SE3STCH_Terr_Det}
as 
\begin{align}
	\dot{\tilde{X}} & =-\mathcal{G}\left(\tilde{\rho}\right)\breve{\overline{\mathbf{Ad}}}_{\hat{\boldsymbol{T}}}\left(\tilde{b}+k_{w}W\right)\label{eq:SE3STCH_Xerr_Det}
\end{align}
with 
\[
\mathcal{G}\left(\tilde{\rho}\right)=\left[\begin{array}{cc}
\frac{\mathbf{I}_{3}+\left[\tilde{\rho}\right]_{\times}+\tilde{\rho}\tilde{\rho}^{\top}}{2} & \mathbf{0}_{3\times3}\\
\mathbf{0}_{3\times3} & \mathcal{R}_{\tilde{\rho}}\left(\tilde{\rho}\right)
\end{array}\right]
\]
and $\mathcal{R}_{\tilde{\rho}}\left(\tilde{\rho}\right)=\tilde{R}\in\mathbb{SO}\left(3\right)$
as given in \eqref{eq:SE3STCH_SO3_Rodr}. Consider the following  potential function 
\begin{equation}
	V\left(\tilde{\rho},\tilde{P},\tilde{b}\right)=\left(\frac{\left\Vert \tilde{\rho}\right\Vert ^{2}}{1+\left\Vert \tilde{\rho}\right\Vert ^{2}}\right)^{2}+2\left\Vert \tilde{P}\right\Vert ^{2}+\frac{1}{2}\tilde{b}^{\top}\Gamma^{-1}\tilde{b}\label{eq:SE3STCH_V_Det}
\end{equation}
for $V:=V\left(\tilde{\rho},\tilde{P},\tilde{b}\right)$ the derivative
of \eqref{eq:SE3STCH_V_Det} is defined by 
\begin{align}
	\dot{V} & =-4\tilde{X}^{\top}\left[\begin{array}{cc}
		\frac{\left\Vert \tilde{\rho}\right\Vert ^{2}}{\left(1+\left\Vert \tilde{\rho}\right\Vert ^{2}\right)^{3}}\mathbf{I}_{3} & \mathbf{0}_{3\times3}\\
		\mathbf{0}_{3\times3} & \mathbf{I}_{3}
	\end{array}\right]\mathcal{G}\left(\tilde{\rho}\right)\breve{\overline{\mathbf{Ad}}}_{\hat{\boldsymbol{T}}}\left(\tilde{b}+k_{w}W\right)-\tilde{b}^{\top}\Gamma^{-1}\dot{\hat{b}}\nonumber \\
	& =-\tilde{X}^{\top}\left[\begin{array}{cc}
		\frac{2\left\Vert \tilde{\rho}\right\Vert ^{2}}{\left(1+\left\Vert \tilde{\rho}\right\Vert ^{2}\right)^{2}}\mathbf{I}_{3} & \mathbf{0}_{3\times3}\\
		\mathbf{0}_{3\times3} & 4\tilde{R}
	\end{array}\right]\breve{\overline{\mathbf{Ad}}}_{\hat{\boldsymbol{T}}}\left(\tilde{b}+k_{w}W\right)-\tilde{b}^{\top}\Gamma^{-1}\dot{\hat{b}}\label{eq:SE3STCH_Vdot_det}
\end{align}
substitute for $\left\Vert \tilde{R}\right\Vert _{I}=\left\Vert \tilde{\rho}\right\Vert ^{2}/\left(1+\left\Vert \tilde{\rho}\right\Vert ^{2}\right)$
and $\boldsymbol{\Upsilon}_{a}\left(\tilde{R}\right)=2\tilde{\rho}/\left(1+\left\Vert \tilde{\rho}\right\Vert ^{2}\right)$
from \eqref{eq:SE3STCH_Ecul_Dist} and \eqref{eq:SE3STCH_VEX_Pa},
respectively, the result in \eqref{eq:SE3STCH_Vdot_det} becomes 
\begin{align}
	\dot{V} & =-\boldsymbol{\Upsilon}\left(\tilde{\boldsymbol{T}}\right)^{\top}\left[\begin{array}{cc}
		\left\Vert \tilde{R}\right\Vert _{I}\mathbf{I}_{3} & \mathbf{0}_{3\times3}\\
		\mathbf{0}_{3\times3} & 4\tilde{R}
	\end{array}\right]\breve{\overline{\mathbf{Ad}}}_{\hat{\boldsymbol{T}}}\left(\tilde{b}+k_{w}W\right)-\tilde{b}^{\top}\Gamma^{-1}\dot{\hat{b}}\label{eq:SE3STCH_Vdot_det1}
\end{align}
such that $\boldsymbol{\Upsilon}\left(\tilde{\boldsymbol{T}}\right)=\left[\boldsymbol{\Upsilon}_{a}^{\top}\left(\tilde{R}\right),\tilde{P}^{\top}\right]^{\top}$,
substituting for $\dot{\hat{b}}$ and $W$ from \eqref{eq:SE3STCH_best_Det}
and \eqref{eq:SE3STCH_Wcorr_Det}, respectively, with $\left\Vert \boldsymbol{\Upsilon}_{a}\left(\tilde{R}\right)\right\Vert ^{2}=4\left(1-\left\Vert \tilde{R}\right\Vert _{I}\right)\left\Vert \tilde{R}\right\Vert _{I}=4\frac{\left\Vert \tilde{\rho}\right\Vert ^{2}}{\left(1+\left\Vert \tilde{\rho}\right\Vert ^{2}\right)^{2}}$
as in \eqref{eq:SE3STCH_VEX2_Pa} yields 
\begin{align}
	\dot{V} & =-k_{w}k_{p}\left\Vert \tilde{R}\right\Vert _{I}\left\Vert \boldsymbol{\Upsilon}_{a}\left(\tilde{\boldsymbol{T}}\right)\right\Vert ^{2}-4k_{w}k_{p}\left(\left\Vert \tilde{R}\right\Vert _{I}^{2}+\left\Vert \tilde{P}\right\Vert ^{2}\right)-k_{b}\left\Vert \tilde{b}\right\Vert ^{2}+k_{b}\tilde{b}^{\top}b\nonumber \\
	& =-4k_{w}k_{p}\frac{\left\Vert \tilde{\rho}\right\Vert ^{4}}{\left(1+\left\Vert \tilde{\rho}\right\Vert ^{2}\right)^{3}}-4k_{w}k_{p}\left(\frac{\left\Vert \tilde{\rho}\right\Vert ^{4}}{\left(1+\left\Vert \tilde{\rho}\right\Vert ^{2}\right)^{2}}+\left\Vert \tilde{P}\right\Vert ^{2}\right)-k_{b}\left\Vert \tilde{b}\right\Vert ^{2}+k_{b}\tilde{b}^{\top}b\label{eq:SE3STCH_Vdot_det2}
\end{align}
applying Young's inequality to $k_{b}\tilde{b}^{\top}b$, one obtains
$k_{b}\tilde{b}^{\top}b\leq\frac{k_{b}}{2}\left\Vert \tilde{b}\right\Vert ^{2}+\frac{k_{b}}{2}\left\Vert b\right\Vert ^{2}$.
Define 
\begin{align*}
	\tilde{Y} & =\left[\frac{\left\Vert \tilde{\rho}\right\Vert ^{2}}{1+\left\Vert \tilde{\rho}\right\Vert ^{2}},\left\Vert \tilde{P}\right\Vert ^{2},\frac{1}{\sqrt{2\gamma}}\tilde{b}^{\top}\right]^{\top}\in\mathbb{R}^{8},\\
	\mathcal{H} & ={\rm diag}\left(4k_{p}k_{w},4k_{p}k_{w},\gamma k_{b}\underline{\mathbf{1}}_{6}^{\top}\right)\in\mathbb{R}^{8\times8}
\end{align*}
therefore, equation \eqref{eq:SE3STCH_Vdot_det2} becomes 
\begin{align}
	\dot{V} & \leq-4k_{w}k_{p}\frac{\left\Vert \tilde{\rho}\right\Vert ^{4}}{\left(1+\left\Vert \tilde{\rho}\right\Vert ^{2}\right)^{3}}-\tilde{Y}^{\top}\mathcal{H}\tilde{Y}+\frac{k_{b}}{2}\left\Vert b\right\Vert ^{2}\nonumber \\
	& \leq-\underline{\lambda}\left(\mathcal{H}\right)V+\frac{k_{b}}{2}\left\Vert b\right\Vert ^{2}\label{eq:SE3STCH_Vdot_det_final}
\end{align}
Let $c_{1}=\underline{\lambda}\left(\mathcal{H}\right)$ and $c_{2}=\frac{k_{b}}{2}\left\Vert b\right\Vert ^{2}$,
thus, the result in \eqref{eq:SE3STCH_Vdot_det_final} implies that
$\hat{X}$ and $\hat{b}$ will eventually converge to the compact
set 
\[
\Xi_{s}=\left\{ \left.\hat{X}\left(t\right),\hat{b}\left(t\right)\right|\lim_{t\rightarrow\infty}\left\Vert \tilde{X}\left(t\right)\right\Vert =\mu_{X},\lim_{t\rightarrow\infty}\left\Vert \tilde{b}\left(t\right)\right\Vert =\mu_{b}\right\} 
\]
with 
\begin{align*}
	\mu_{X} & =\sqrt{\frac{c_{2}}{c_{1}}},\hspace{1em}\mu_{b}=\sqrt{\frac{2c_{2}}{c_{1}\gamma}}
\end{align*}
and 
\begin{align*}
	\left\Vert \tilde{X}\left(t\right)\right\Vert  & \leq\sqrt{\left(V\left(0\right)-\frac{c_{2}}{c_{1}}\right)\exp\left(-c_{1}t\right)+\frac{c_{2}}{c_{1}}}\\
	\left\Vert \tilde{b}\left(t\right)\right\Vert  & \leq\frac{1}{\gamma}\sqrt{\left(V\left(0\right)-\frac{c_{2}}{c_{1}}\right)\exp\left(-c_{1}t\right)+\frac{c_{2}}{c_{1}}}
\end{align*}
The result obtained in \eqref{eq:SE3STCH_Vdot_det_final} is similar
to Lemma 1.2 in \cite{ge2004adaptive} which confirms the result in
Theorem \ref{thm:SE3STCH_1}. Theorem \ref{thm:SE3STCH_1} is developed
for deterministic observers, assuming absence of noises in the system
dynamics. Hence, Lyapunov's direct method guarantees that for ${\rm Tr}\left\{ \tilde{R}\left(0\right)\right\} \neq-1$,
$\boldsymbol{\Upsilon}\left(\tilde{\boldsymbol{T}}\right)$ converges
to a small neighborhood of the origin. However, if the velocity vector
$\mathcal{Y}_{m}$ is contaminated with noise such that $\left(\omega\neq0\right)$,
it would no longer be convenient to express the derivative of \eqref{eq:SE3STCH_V_Det}
similar to \eqref{eq:SE3STCH_Vdot_det}. Therefore, the derivative
of \eqref{eq:SE3STCH_V_Det} should be expressed analogously to the
differential operator in Definition \ref{def:SE3STCH_2} and consequently,
the covariance matrix $\mathcal{Q}^{2}$ appears there. As a result,
one solution is to reformulate the potential function in \eqref{eq:SE3STCH_V_Det}
such that $\tilde{\rho}$ and $\tilde{P}$ are of order higher than
two \cite{deng2001stabilization,deng1997stochastic}. Clearly, this
is not the case in Theorem \ref{thm:SE3STCH_1} as well as in previous
studies such as \cite{rehbinder2003pose,baldwin2007complementary,baldwin2009nonlinear,hua2011observer,vasconcelos2010nonlinear}.

\subsection{Nonlinear Stochastic Pose Filter in Stratonovich Sense}

Generally, nonlinear deterministic attitude or attitude-position filters
assume that velocity measurements are subject only to constant bias
(for example \cite{crassidis2007survey,mahony2008nonlinear,rehbinder2003pose,baldwin2007complementary,baldwin2009nonlinear,hua2011observer}).
In contrast, the velocity vector $\mathcal{Y}_{m}$ is contaminated
not only with bias but also noise components. The added components
could impair the estimation process of the true position and attitude.
As such, the aim is to design a nonlinear stochastic filter evolved
directly on $\mathbb{SE}\left(3\right)$ in the sense of Stratonovich
\cite{stratonovich1967topics} considering that measurement in the
velocity vector $\mathcal{Y}_{m}$ is contaminated with constant bias
and a wide-band of Gaussian random noise with zero mean. Stochastic
differential equations can be defined and solved in the sense of Ito's
integral \cite{ito1984lectures}. Alternatively, Stratonovich's integral
\cite{stratonovich1967topics} can be employed for solving stochastic
differential equations. The common feature between Stratonovich and
Ito integral is that if the associated function multiplied by $d\beta$
is continuous and Lipschitz, the mean square limit exists. The Ito
integral is defined for functional on $\left\{ \beta\left(\tau\right),\tau\leq t\right\} $
which is more natural but it does not obey the chain rule. Conversely,
Stratonovich is a well-defined Riemann integral for the sampled function,
it has a continuous partial derivative with respect to $\beta$, it
obeys the chain rule, and it is more convenient for colored noise
\cite{stratonovich1967topics,jazwinski2007stochastic}. Hence, the
Stratonovich integral is defined for explicit functions of $\beta$.
In case of a wide-band of random colored noise process being attached
to the velocity measurements, for $X=\left[\rho^{\top},P^{\top}\right]^{\top}$
with $X\left(t_{0}\right)=0$, the solution of \eqref{eq:SE3STCH_X_Ito}
is defined by 
\begin{equation}
	X\left(t\right)=\int_{t_{0}}^{t}f\left(\rho\left(\tau\right),b\left(\tau\right)\right)d\tau+\int_{t_{0}}^{t}\mathcal{G}\left(\rho\left(\tau\right)\right)\mathcal{Q}d\beta\label{eq:SE3STCH_solution}
\end{equation}
if the problem has been considered and solved directly in the sense
of Ito, the expected value of \eqref{eq:SE3STCH_solution} is 
\[
\mathbb{E}\left[X\right]\neq\int_{t_{0}}^{t}\mathbb{E}\left[f\left(\rho\left(\tau\right),b\left(\tau\right)\right)\right]d\tau
\]
Hence, Stratonovich came up with the Wong-Zakai correction factor
to balance any colored noise that may be introduced to the system
dynamics and to end with $\mathbb{E}\left[X\right]=\int_{t_{0}}^{t}\mathbb{E}\left[f\left(\rho,b\right)\right]d\tau$.
A remarkable advantage of Stratonovich is its applicability to white
noise as well as colored noise which makes the filter more robust
for real time applications \cite{stratonovich1967topics,khasminskii1980stochastic,jazwinski2007stochastic}.
Let us assume that the attitude dynamics in \eqref{eq:SE3STCH_X_Ito}
were defined in the sense of Stratonovich \cite{stratonovich1967topics}.
Therefore, the equivalent Ito \cite{khasminskii1980stochastic,jazwinski2007stochastic,ito1984lectures}
can be expressed as 
\begin{align}
	\left[dX\right]_{i}= & \left[f\left(\rho,b\right)\right]_{i}dt+\sum_{k=1}^{6}\sum_{j=1}^{6}\frac{\mathcal{Q}_{j,j}^{2}}{2}\mathcal{G}_{kj}\left(\rho\right)\frac{\partial\mathcal{G}_{ij}\left(\rho\right)}{\partial X_{k}}dt+\left[\mathcal{G}\left(\rho\right)\mathcal{Q}d\beta\right]_{i}\label{eq:SE3STCH_Stoch_strat_i}
\end{align}
where both $f\left(\rho,b\right)$ and $\mathcal{G}\left(\rho\right)$
are defined in \eqref{eq:SE3STCH_X_Ito}. $\sum_{k=1}^{6}\sum_{j=1}^{6}\frac{\mathcal{Q}_{j,j}^{2}}{2}\mathcal{G}_{kj}\left(\rho\right)\frac{\partial\mathcal{G}_{ij}\left(\rho\right)}{\partial\rho_{k}}$
is termed the Wong-Zakai correction factor of stochastic differential
equations (SDEs) in the sense of Ito \cite{wong1965convergence},
and $i,j,k=1,\ldots,6$ denote $i$th, $j$th and/or $k$th elements
of the associated vector or matrix. Assume that $\boldsymbol{\mathcal{W}}\left(\rho\right)=\left[\boldsymbol{\mathcal{W}}_{\rho}^{\top},\boldsymbol{\mathcal{W}}_{P}^{\top}\right]^{\top}\in\mathbb{R}^{6}$.
Let $\boldsymbol{\mathcal{W}}_{\rho i}=\sum_{k=1}^{3}\sum_{j=1}^{3}\frac{\mathcal{Q}_{j,j}^{2}}{2}\mathcal{G}_{kj}\left(\rho\right)\frac{\partial\mathcal{G}_{ij}\left(\rho\right)}{\partial\rho_{k}}$,
therefore, for $i=1$ 
\begin{align*}
	\boldsymbol{\mathcal{W}}_{\rho i}= & \frac{1}{4}\left(\left(1+\rho_{1}^{2}\right)\rho_{1}\mathcal{Q}_{1,1}^{2}+\left(\rho_{1}\rho_{2}-\rho_{3}\right)\rho_{2}\mathcal{Q}_{2,2}^{2}+\left(\rho_{2}+\rho_{1}\rho_{3}\right)\rho_{3}\mathcal{Q}_{3,3}^{2}\right)
\end{align*}
Thus, one can find that for $i=1,2,3$, $\boldsymbol{\mathcal{W}}_{\rho}\in\mathbb{R}^{3}$
can be defined after some steps of calculations as follows 
\begin{align}
	\boldsymbol{\mathcal{W}}_{\rho} & =\frac{1}{4}\left(\mathbf{I}_{3}+\left[\rho\right]_{\times}+\rho\rho^{\top}\right)\mathcal{Q}_{\Omega}^{2}\rho\label{eq:SE3STCH_WONG-Zaki}
\end{align}
And $\boldsymbol{\mathcal{W}}_{Pi}=\sum_{k=4}^{6}\sum_{j=4}^{6}\frac{\mathcal{Q}_{j,j}^{2}}{2}\mathcal{G}_{kj}\left(\rho\right)\frac{\partial\mathcal{G}_{ij}\left(\rho\right)}{\partial P_{k}}=0$,
for $i=4,5,6$. This implies that 
\begin{align}
	\boldsymbol{\mathcal{W}}_{P} & =\underline{\mathbf{0}}_{3}\in\mathbb{R}^{3}\label{eq:SE3STCH_WONG-Zaki2}
\end{align}
Manipulating equations \eqref{eq:SE3STCH_Stoch_strat_i}, \eqref{eq:SE3STCH_WONG-Zaki}
and \eqref{eq:SE3STCH_WONG-Zaki2}, the stochastic dynamics of the
Rodriguez vector can be expressed as 
\begin{align}
	dX= & \left(f\left(\rho,b\right)\left(\mathcal{Y}_{m}-b\right)+\boldsymbol{\mathcal{W}}\left(\rho\right)\right)dt-\mathcal{G}\left(\rho\right)\mathcal{Q}d\beta\label{eq:SE3STCH_Stoch_strat}
\end{align}
Assume that the elements of covariance matrix $\mathcal{Q}^{2}$ are
upper bounded by $\sigma$ as given in \eqref{eq:SE3STCH_g_factor}
such that the bound of $\sigma$ is unknown for nonnegative elements.
Consider the nonlinear stochastic pose filter design 
\begin{align}
	\dot{\hat{\boldsymbol{T}}}= & \hat{\boldsymbol{T}}\left[\mathcal{Y}_{m}-\hat{b}+k_{w}W+\breve{\overline{\mathbf{Ad}}}_{\hat{\boldsymbol{T}}^{-1}}\left[\begin{array}{cc}
		\frac{1}{2}\frac{1}{1-\left\Vert \tilde{R}\right\Vert _{I}}\mathbf{I}_{3} & \mathbf{0}_{3\times3}\\
		\mathbf{0}_{3\times3} & \mathbf{0}_{3\times3}
	\end{array}\right]{\rm diag}\left(\boldsymbol{\Upsilon}\left(\tilde{\boldsymbol{T}}\right)\right)\hat{\sigma}\right]_{\wedge},\hspace{1em}\hat{\boldsymbol{T}}\left(0\right)\in\mathbb{SE}\left(3\right)\label{eq:SE3STCH_dTest_Strat}\\
	\dot{\hat{b}}= & -\Gamma\breve{\overline{\mathbf{Ad}}}_{\hat{\boldsymbol{T}}}^{\top}\left[\begin{array}{cc}
		\left\Vert \tilde{R}\right\Vert _{I}\mathbf{I}_{3} & \mathbf{0}_{3\times3}\\
		\mathbf{0}_{3\times3} & 4\left\Vert \tilde{P}\right\Vert ^{2}\tilde{R}^{\top}
	\end{array}\right]\boldsymbol{\Upsilon}\left(\tilde{\boldsymbol{T}}\right)-k_{b}\Gamma\hat{b}\label{eq:SE3STCH_best_Strat}\\
	\dot{\hat{\sigma}}= & \Pi\left(\frac{1}{4}\frac{\left\Vert \tilde{R}\right\Vert _{I}}{1-\left\Vert \tilde{R}\right\Vert _{I}}{\rm diag}\left(\left[\begin{array}{c}
		\boldsymbol{\Upsilon}_{a}\left(\tilde{R}\right)\\
		\underline{\mathbf{0}}_{3}
	\end{array}\right]\right)+k_{w}k_{p}\left[\begin{array}{cc}
		\left\Vert \tilde{R}\right\Vert _{I}\mathcal{D}_{\Upsilon}^{\top} & \mathbf{0}_{3\times3}\\
		\mathbf{0}_{3\times3} & \mathbf{0}_{3\times3}
	\end{array}\right]\right)\boldsymbol{\Upsilon}\left(\tilde{\boldsymbol{T}}\right)-k_{\sigma}\Pi\hat{\sigma}\label{eq:SE3STCH_gest_strat}\\
	W= & k_{p}\breve{\overline{\mathbf{Ad}}}_{\hat{\boldsymbol{T}}^{-1}}\left(\frac{1}{\varepsilon}\left[\begin{array}{cc}
		\frac{2-\left\Vert \tilde{R}\right\Vert _{I}}{1-\left\Vert \tilde{R}\right\Vert _{I}}\mathbf{I}_{3} & \mathbf{0}_{3\times3}\\
		\mathbf{0}_{3\times3} & \tilde{R}^{\top}
	\end{array}\right]\boldsymbol{\Upsilon}\left(\tilde{\boldsymbol{T}}\right)+\left[\begin{array}{cc}
		\mathcal{D}_{\Upsilon} & \mathbf{0}_{3\times3}\\
		\mathbf{0}_{3\times3} & \mathbf{0}_{3\times3}
	\end{array}\right]\hat{\sigma}\right)\label{eq:SE3STCH_Wcorr_Strat}
\end{align}
where $\mathcal{Y}_{m}=\left[\Omega_{m}^{\top},V_{m}^{\top}\right]^{\top}$
denotes the measured vector of angular and translational velocity
defined in \eqref{eq:SE3STCH_Angular} and \eqref{eq:SE3STCH_V_Trans},
respectively. $\hat{b}=\left[\hat{b}_{\Omega}^{\top},\hat{b}_{V}^{\top}\right]^{\top}\in\mathbb{R}^{6}$
and $\hat{\sigma}=\left[\hat{\sigma}_{\Omega}^{\top},\hat{\sigma}_{V}^{\top}\right]^{\top}\in\mathbb{R}^{6}$
are estimates of the unknown parameter $b$ and $\sigma$, respectively,
$\tilde{\boldsymbol{T}}=\boldsymbol{T}_{y}\hat{\boldsymbol{T}}^{-1}$,
$\boldsymbol{\Upsilon}\left(\tilde{\boldsymbol{T}}\right)=\left[\boldsymbol{\Upsilon}_{a}^{\top}\left(\tilde{R}\right),\tilde{P}^{\top}\right]^{\top}$
as in \eqref{eq:SE3STCH_Vex}, $\boldsymbol{\Upsilon}_{a}\left(\tilde{R}\right)=\mathbf{vex}\left(\boldsymbol{\mathcal{P}}_{a}\left(\tilde{R}\right)\right)$
as given in \eqref{eq:SE3STCH_VEX_Pa}, $\left\Vert \tilde{R}\right\Vert _{I}=\frac{1}{4}{\rm Tr}\left\{ \mathbf{I}_{3}-\tilde{R}\right\} $
is the Euclidean distance of $\tilde{R}$ as defined in \eqref{eq:SE3STCH_Ecul_Dist},
and $\mathcal{D}_{\Upsilon}=\left[\boldsymbol{\Upsilon}_{a}\left(\tilde{R}\right),\boldsymbol{\Upsilon}_{a}\left(\tilde{R}\right),\boldsymbol{\Upsilon}_{a}\left(\tilde{R}\right)\right]$.
Also, $\breve{\overline{\mathbf{Ad}}}_{\hat{\boldsymbol{T}}}=\left[\begin{array}{cc}
\hat{R} & \mathbf{0}_{3\times3}\\
\left[\hat{P}\right]_{\times}\hat{R} & \hat{R}
\end{array}\right]$, $\Gamma=\left[\begin{array}{cc}
\Gamma_{\Omega} & \mathbf{0}_{3\times3}\\
\mathbf{0}_{3\times3} & \Gamma_{V}
\end{array}\right]=\gamma\mathbf{I}_{6}$, and $\Pi=\left[\begin{array}{cc}
\Pi_{\Omega} & \mathbf{0}_{3\times3}\\
\mathbf{0}_{3\times3} & \Pi_{V}
\end{array}\right]=\bar{\pi}\mathbf{I}_{6}$ are adaptation gains with $\Gamma_{\Omega},\Gamma_{V},\Pi_{\Omega},\Pi_{V}\in\mathbb{R}^{3\times3}$
where $\gamma,\bar{\pi}>0$, $\varepsilon>0$ is a small constant,
and $k_{b}$, $k_{\sigma}$, $k_{p}$ and $k_{w}$ are positive constants. 
\begin{thm}
	\label{thm:SE3STCH_2}\textbf{ }Consider the homogeneous transformation
	matrix dynamics in \eqref{eq:SE3STCH_T_Dynamics} with velocity measurements
	$\mathcal{Y}_{m}=\left[\Omega_{m}^{\top},V_{m}^{\top}\right]^{\top}$
	in \eqref{eq:SE3STCH_Angular} and \eqref{eq:SE3STCH_V_Trans}. Let
	Assumption \ref{Assum:SE3STCH_1} hold and assume that the vector
	measurements in \eqref{eq:SE3STCH_Vect_R} are normalized to \eqref{eq:SE3STCH_Vector_norm}.
	Let $\boldsymbol{T}_{y}$ be reconstructed using the vector measurements
	in \eqref{eq:SE3STCH_Vec_Landmark} and \eqref{eq:SE3STCH_Vector_norm},
	and be coupled with the observer in \eqref{eq:SE3STCH_dTest_Strat},
	\eqref{eq:SE3STCH_best_Strat}, \eqref{eq:SE3STCH_gest_strat} and
	\eqref{eq:SE3STCH_Wcorr_Strat}. Assume the design parameters $\Gamma$,
	$\Pi$, $\varepsilon$, $k_{b}$, $k_{\sigma}$, $k_{p}$ and $k_{w}$
	are chosen appropriately with $\varepsilon$ being selected sufficiently
	small. When velocity measurements $\mathcal{Y}_{m}$ are contaminated
	with bias and noise $\left(\omega\neq0\right)$, $\tilde{X}\left(0\right)=\left[\tilde{\rho}\left(0\right)^{\top},\tilde{P}\left(0\right)^{\top}\right]^{\top}\in\mathbb{R}^{6}$,
	and $\tilde{X}\left(0\right)\neq\underline{\mathbf{0}}_{6}$, then
	1) the errors $\left(\tilde{\boldsymbol{T}},\tilde{b},\tilde{\sigma}\right)$
	are regulated to the neighborhood of the equilibrium set $\mathcal{S}=\left\{ \left(\tilde{\boldsymbol{T}},\tilde{b},\tilde{\sigma}\right)\in\mathbb{SE}\left(3\right)\times\mathbb{R}^{6}\times\mathbb{R}^{6}:\tilde{\boldsymbol{T}}=\mathbf{I}_{4},\tilde{b}=\underline{\mathbf{0}}_{6},\tilde{\sigma}=\underline{\mathbf{0}}_{6}\right\} $;
	and 2) $\left[\tilde{X}^{\top},\tilde{b}^{\top},\tilde{\sigma}^{\top}\right]^{\top}$is
	semi-globally uniformly ultimately bounded in mean square. 
\end{thm}
\textbf{Proof: }Let the error in the homogeneous transformation matrix
$\boldsymbol{T}$ be given as in \eqref{eq:SE3STCH_T_error} and the
error in vector $b$ be defined as in \eqref{eq:SE3STCH_b_tilde}.
Therefore, the derivative of homogeneous transformation matrix error
$\tilde{\boldsymbol{T}}$ in \eqref{eq:SE3STCH_T_error} in incremental
form can be obtained from \eqref{eq:SE3STCH_T_Dynam_Noise} and \eqref{eq:SE3STCH_dTest_Strat}
by 
\begin{align}
	d\tilde{\boldsymbol{T}}= & d\boldsymbol{T}\hat{\boldsymbol{T}}^{-1}+\boldsymbol{T}d\hat{\boldsymbol{T}}^{-1}\nonumber \\
	= & \boldsymbol{T}\left[\mathcal{Y}_{m}-b\right]_{\wedge}\hat{\boldsymbol{T}}^{-1}dt-\boldsymbol{T}\left[\mathcal{Q}d\beta\right]_{\wedge}\hat{\boldsymbol{T}}^{-1}\nonumber \\
	& -\boldsymbol{T}\left[\mathcal{Y}_{m}-\hat{b}+k_{w}W+\breve{\overline{\mathbf{Ad}}}_{\hat{\boldsymbol{T}}^{-1}}\left[\begin{array}{cc}
		\frac{1}{2}\frac{1}{1-\left\Vert \tilde{R}\right\Vert _{I}}\mathbf{I}_{3} & \mathbf{0}_{3\times3}\\
		\mathbf{0}_{3\times3} & \mathbf{0}_{3\times3}
	\end{array}\right]{\rm diag}\left(\boldsymbol{\Upsilon}\left(\tilde{\boldsymbol{T}}\right)\right)\hat{\sigma}\right]_{\wedge}\hat{\boldsymbol{T}}^{-1}dt\nonumber \\
	= & -\tilde{\boldsymbol{T}}\hat{\boldsymbol{T}}\left[\tilde{b}+k_{w}W+\breve{\overline{\mathbf{Ad}}}_{\hat{\boldsymbol{T}}^{-1}}\left[\begin{array}{cc}
		\frac{1}{2}\frac{1}{1-\left\Vert \tilde{R}\right\Vert _{I}}\mathbf{I}_{3} & \mathbf{0}_{3\times3}\\
		\mathbf{0}_{3\times3} & \mathbf{0}_{3\times3}
	\end{array}\right]{\rm diag}\left(\boldsymbol{\Upsilon}\left(\tilde{\boldsymbol{T}}\right)\right)\hat{\sigma}\right]_{\wedge}\hat{\boldsymbol{T}}^{-1}dt-\tilde{\boldsymbol{T}}\hat{\boldsymbol{T}}\left[\mathcal{Q}d\beta\right]_{\wedge}\hat{\boldsymbol{T}}^{-1}\nonumber \\
	= & -\tilde{\boldsymbol{T}}\left[\breve{\overline{\mathbf{Ad}}}_{\hat{\boldsymbol{T}}}\left(\tilde{b}+k_{w}W\right)+\left[\begin{array}{cc}
		\frac{1}{2}\frac{1}{1-\left\Vert \tilde{R}\right\Vert _{I}}\mathbf{I}_{3} & \mathbf{0}_{3\times3}\\
		\mathbf{0}_{3\times3} & \mathbf{0}_{3\times3}
	\end{array}\right]{\rm diag}\left(\boldsymbol{\Upsilon}\left(\tilde{\boldsymbol{T}}\right)\right)\hat{\sigma}\right]_{\wedge}dt-\tilde{\boldsymbol{T}}\left[\breve{\overline{\mathbf{Ad}}}_{\hat{\boldsymbol{T}}}\mathcal{Q}d\beta\right]_{\wedge}\label{eq:SE3STCH_dT_tilde_ito}
\end{align}
where $\dot{\hat{\boldsymbol{T}}}^{-1}=-\hat{\boldsymbol{T}}^{-1}\dot{\hat{\boldsymbol{T}}}\hat{\boldsymbol{T}}^{-1}$,
and $\tilde{b}=\left[\tilde{b}_{\Omega}^{\top},\tilde{b}_{V}^{\top}\right]^{\top}$.
Considering the math identity in \eqref{eq:SE3STCH_Identity5} we
have $\hat{\boldsymbol{T}}\left[\tilde{b}\right]_{\wedge}\hat{\boldsymbol{T}}^{-1}=\left[\breve{\overline{\mathbf{Ad}}}_{\hat{\boldsymbol{T}}}\tilde{b}\right]_{\wedge}$,
and from the math identity in \eqref{eq:SE3STCH_Identity3} and \eqref{eq:SE3STCH_Identity4},
we have $\breve{\overline{\mathbf{Ad}}}_{\hat{\boldsymbol{T}}}\breve{\overline{\mathbf{Ad}}}_{\hat{\boldsymbol{T}}^{-1}}=\mathbf{I}_{6}$.
Similarly to transition from \eqref{eq:SE3STCH_T_Ito} to \eqref{eq:SE3STCH_X_Ito},
extraction of vector dynamics in \eqref{eq:SE3STCH_dT_tilde_ito}
can be expressed as \eqref{eq:SE3STCH_StochErr_itop} and \eqref{eq:SE3STCH_StochErr_ito}
in Stratonovich's representation \cite{stratonovich1967topics} as
follows 
\begin{align}
	d\tilde{X}= & -\left[\begin{array}{cc}
		\frac{\mathbf{I}_{3}+\left[\tilde{\rho}\right]_{\times}+\tilde{\rho}\tilde{\rho}^{\top}}{2} & \mathbf{0}_{3\times3}\\
		\mathbf{0}_{3\times3} & \mathcal{R}_{\tilde{\rho}}\left(\tilde{\rho}\right)
	\end{array}\right]\left(\breve{\overline{\mathbf{Ad}}}_{\hat{\boldsymbol{T}}}\left(\tilde{b}+k_{w}W\right)+\left[\begin{array}{cc}
		\frac{1}{2}\frac{1}{1-\left\Vert \tilde{R}\right\Vert _{I}}\mathbf{I}_{3} & \mathbf{0}_{3\times3}\\
		\mathbf{0}_{3\times3} & \mathbf{0}_{3\times3}
	\end{array}\right]{\rm diag}\left(\boldsymbol{\Upsilon}\left(\tilde{\boldsymbol{T}}\right)\right)\hat{\sigma}\right)dt\nonumber \\
	& -\left[\begin{array}{cc}
		\frac{\mathbf{I}_{3}+\left[\tilde{\rho}\right]_{\times}+\tilde{\rho}\tilde{\rho}^{\top}}{2} & \mathbf{0}_{3\times3}\\
		\mathbf{0}_{3\times3} & \mathcal{R}_{\tilde{\rho}}\left(\tilde{\rho}\right)
	\end{array}\right]\breve{\overline{\mathbf{Ad}}}_{\hat{\boldsymbol{T}}}\mathcal{Q}d\beta\label{eq:SE3STCH_StochErr_itop}
\end{align}
Or more simply as 
\begin{align}
	d\tilde{X}= & -f_{\tilde{X}}dt-\mathcal{G}\left(\tilde{\rho}\right)\breve{\overline{\mathbf{Ad}}}_{\hat{\boldsymbol{T}}}\mathcal{Q}d\beta\label{eq:SE3STCH_StochErr_ito}
\end{align}
where 
\begin{align*}
	\mathcal{G}\left(\tilde{\rho}\right) & =\left[\begin{array}{cc}
		g_{\tilde{\rho}}\left(\tilde{\rho}\right) & \mathbf{0}_{3\times3}\\
		\mathbf{0}_{3\times3} & g_{\tilde{P}}\left(\tilde{\rho}\right)
	\end{array}\right]\\
	g_{\tilde{\rho}}\left(\tilde{\rho}\right) & =\frac{\mathbf{I}_{3}+\left[\tilde{\rho}\right]_{\times}+\tilde{\rho}\tilde{\rho}^{\top}}{2}\\
	g_{\tilde{P}}\left(\tilde{\rho}\right) & =\mathcal{R}_{\tilde{\rho}}\left(\tilde{\rho}\right)
\end{align*}
and 
\[
f_{\tilde{X}}=-\mathcal{G}\left(\tilde{\rho}\right)\left(\breve{\overline{\mathbf{Ad}}}_{\hat{\boldsymbol{T}}}\left(\tilde{b}+k_{w}W\right)+\left[\begin{array}{cc}
\frac{1}{2}\frac{1}{1-\left\Vert \tilde{R}\right\Vert _{I}}\mathbf{I}_{3} & \mathbf{0}_{3\times3}\\
\mathbf{0}_{3\times3} & \mathbf{0}_{3\times3}
\end{array}\right]{\rm diag}\left(\boldsymbol{\Upsilon}\left(\tilde{\boldsymbol{T}}\right)\right)\hat{\sigma}\right)
\]
One can re-define 
\begin{align*}
	\bar{\omega}_{\Omega} & =\hat{R}\omega_{\Omega}\\
	\bar{\omega}_{V} & =\left[\hat{P}\right]_{\times}\hat{R}\omega_{\Omega}+\hat{R}\omega_{V}
\end{align*}
for all $\bar{\omega}_{\Omega},\bar{\omega}_{V}\in\mathbb{R}^{3}$
such that 
\[
\bar{\omega}_{\Omega}=\bar{\mathcal{Q}}_{\Omega}\frac{d\bar{\beta}_{\Omega}}{dt},\hspace{1em}\bar{\omega}_{V}=\bar{\mathcal{Q}}_{V}\frac{d\bar{\beta}_{V}}{dt}
\]
with 
\begin{align*}
	\bar{\beta} & =\left[\bar{\beta}_{\Omega}^{\top},\bar{\beta}_{V}^{\top}\right]^{\top}\in\mathbb{R}^{6}\\
	\bar{\mathcal{Q}} & =\left[\begin{array}{cc}
		\bar{\mathcal{Q}}_{\Omega} & \mathbf{0}_{3\times3}\\
		\mathbf{0}_{3\times3} & \bar{\mathcal{Q}}_{V}
	\end{array}\right]\in\mathbb{R}^{6\times6}
\end{align*}
Thus, the dynamics in \eqref{eq:SE3STCH_dT_tilde_ito} and \eqref{eq:SE3STCH_StochErr_ito}
can be re-expressed, respectively, as 
\begin{align}
	d\tilde{\boldsymbol{T}} & =-\tilde{\boldsymbol{T}}\left[\breve{\overline{\mathbf{Ad}}}_{\hat{\boldsymbol{T}}}\left(\tilde{b}+k_{w}W\right)+\left[\begin{array}{cc}
		\frac{1}{2}\frac{1}{1-\left\Vert \tilde{R}\right\Vert _{I}}\mathbf{I}_{3} & \mathbf{0}_{3\times3}\\
		\mathbf{0}_{3\times3} & \mathbf{0}_{3\times3}
	\end{array}\right]{\rm diag}\left(\boldsymbol{\Upsilon}\left(\tilde{\boldsymbol{T}}\right)\right)\hat{\sigma}\right]_{\wedge}dt-\tilde{\boldsymbol{T}}\left[\bar{\mathcal{Q}}d\bar{\beta}\right]_{\wedge}\label{eq:SE3STCH_StochErr_itoT1}
\end{align}
\begin{align}
	d\tilde{X}= & -f_{\tilde{X}}dt-\mathcal{G}\left(\tilde{\rho}\right)\bar{\mathcal{Q}}d\bar{\beta}\label{eq:SE3STCH_StochErr_ito1-1}
\end{align}
Hence, in view of \eqref{eq:SE3STCH_Stoch_strat_i} and \eqref{eq:SE3STCH_Stoch_strat},
the error dynamics in \eqref{eq:SE3STCH_StochErr_ito1-1} can be re-expressed
in the sense of Ito \cite{ito1984lectures, hashim2018Stoch} as 
\begin{align}
	d\tilde{X}= & -\mathcal{G}\left(\tilde{\rho}\right)\left(\breve{\overline{\mathbf{Ad}}}_{\hat{\boldsymbol{T}}}\left(\tilde{b}+k_{w}W\right)+\left[\begin{array}{cc}
		\frac{1}{2}\frac{1}{1-\left\Vert \tilde{R}\right\Vert _{I}}\mathbf{I}_{3} & \mathbf{0}_{3\times3}\\
		\mathbf{0}_{3\times3} & \mathbf{0}_{3\times3}
	\end{array}\right]{\rm diag}\left(\boldsymbol{\Upsilon}\left(\tilde{\boldsymbol{T}}\right)\right)\hat{\sigma}\right)dt+\left[\begin{array}{c}
		\boldsymbol{\mathcal{W}}_{\tilde{\rho}}\\
		\boldsymbol{\mathcal{W}}_{\tilde{P}}
	\end{array}\right]dt-\mathcal{G}\left(\tilde{\rho}\right)\bar{\mathcal{Q}}d\bar{\beta}\label{eq:SE3STCH_StochErr_Strat1}
\end{align}
with $\boldsymbol{\mathcal{W}}_{\tilde{\rho}}=\frac{1}{4}\left(\mathbf{I}_{3}+\left[\tilde{\rho}\right]_{\times}+\tilde{\rho}\tilde{\rho}^{\top}\right)\mathcal{\bar{Q}}_{\Omega}^{2}\tilde{\rho}$
and $\boldsymbol{\mathcal{W}}_{\tilde{P}}=\underline{\mathbf{0}}_{3}$
as defined in \eqref{eq:SE3STCH_WONG-Zaki} and \eqref{eq:SE3STCH_WONG-Zaki2},
respectively, which can be further simplified as shown below 
\begin{align}
	d\tilde{X}= & \left(-f_{\tilde{X}}+\boldsymbol{\mathcal{W}}\left(\tilde{\rho}\right)\right)dt-\mathcal{G}\left(\tilde{\rho}\right)\bar{\mathcal{Q}}d\bar{\beta}\nonumber \\
	= & \mathcal{F}dt-\mathcal{G}\left(\tilde{\rho}\right)\bar{\mathcal{Q}}d\bar{\beta}\label{eq:SE3STCH_StochErr_Strat}
\end{align}
where $\mathcal{F}=\left[\mathcal{F}_{\tilde{\rho}}^{\top},\mathcal{F}_{\tilde{P}}^{\top}\right]^{\top}=-f_{\tilde{X}}+\boldsymbol{\mathcal{W}}\left(\tilde{\rho}\right)$.
Let us re-define $\sigma$ as the upper bound of $\bar{\mathcal{Q}}^{2}$
with $\sigma=\left[\sigma_{\Omega}^{\top},\sigma_{V}^{\top}\right]^{\top}\in\mathbb{R}^{6}$
and $\sigma_{\Omega},\sigma_{V}\in\mathbb{R}^{3}$ such that 
\begin{equation}
	\sigma=\left[{\rm max}\left\{ \bar{\mathcal{Q}}_{\left(1,1\right)}^{2}\right\} ,{\rm max}\left\{ \bar{\mathcal{Q}}_{\left(2,2\right)}^{2}\right\} ,\ldots,{\rm max}\left\{ \mathcal{\bar{Q}}_{\left(6,6\right)}^{2}\right\} \right]^{\top}\label{eq:SE3STCH_g_factor-1}
\end{equation}
Let the error in $\sigma$ be defined similar to \eqref{eq:SE3STCH_sigma_tilde_strat}
with $\tilde{\sigma}=\sigma-\hat{\sigma}$. Consider the following potential function 
\begin{align}
	V\left(\tilde{\rho},\tilde{P},\tilde{b},\tilde{\sigma}\right)= & \left(\frac{\left\Vert \tilde{\rho}\right\Vert ^{2}}{1+\left\Vert \tilde{\rho}\right\Vert ^{2}}\right)^{2}+\left\Vert \tilde{P}\right\Vert ^{4}+\frac{1}{2}\tilde{b}^{\top}\Gamma^{-1}\tilde{b}+\frac{1}{2}\tilde{\sigma}^{\top}\Pi^{-1}\tilde{\sigma}\label{eq:SE3STCH_LyapV_strat}
\end{align}
For $V:=V\left(\tilde{\rho},\tilde{P},\tilde{b},\tilde{\sigma}\right)$,
the differential operator $\mathcal{L}V$ in Definition \ref{def:SE3STCH_2}
can be written as 
\begin{align}
	\mathcal{L}V & =V_{\tilde{\rho}}^{\top}\mathcal{F}_{\tilde{\rho}}+\frac{1}{2}{\rm Tr}\left\{ g_{\tilde{\rho}}^{\top}V_{\tilde{\rho}\tilde{\rho}}g_{\tilde{\rho}}\mathcal{\bar{Q}}_{\Omega}^{2}\right\} +V_{\tilde{P}}^{\top}\mathcal{F}_{\tilde{P}}+\frac{1}{2}{\rm Tr}\left\{ g_{\tilde{P}}^{\top}V_{\tilde{P}\tilde{P}}g_{\tilde{P}}\bar{\mathcal{Q}_{V}}^{2}\right\} -\tilde{b}^{\top}\Gamma^{-1}\dot{\hat{b}}-\tilde{\sigma}^{\top}\Pi^{-1}\dot{\hat{\sigma}}\label{eq:SE3STCH_LV_strat}
\end{align}
One can easily show that the first and second partial derivatives
of \eqref{eq:SE3STCH_LyapV_strat} in terms of $\tilde{\rho}$ can
be obtained as follows 
\begin{align}
	V_{\tilde{\rho}}= & 4\frac{\left\Vert \tilde{\rho}\right\Vert ^{2}}{\left(1+\left\Vert \tilde{\rho}\right\Vert ^{2}\right)^{3}}\tilde{\rho}\label{eq:SE3STCH_LyapVv_ito}\\
	V_{\tilde{\rho}\tilde{\rho}}= & 4\frac{\left(1+\left\Vert \tilde{\rho}\right\Vert ^{2}\right)\left\Vert \tilde{\rho}\right\Vert ^{2}\mathbf{I}_{3}+\left(2-4\left\Vert \tilde{\rho}\right\Vert ^{2}\right)\tilde{\rho}\tilde{\rho}^{\top}}{\left(1+\left\Vert \tilde{\rho}\right\Vert ^{2}\right)^{4}}\label{eq:SE3STCH_LyapVvv_ito}
\end{align}
\textbf{ }Similarly, the first and second partial derivatives of \eqref{eq:SE3STCH_LyapV_strat}
in terms of $\tilde{P}$ can be obtained as follows 
\begin{align}
	V_{\tilde{P}} & =4\left\Vert \tilde{P}\right\Vert ^{2}\tilde{P}\label{eq:SE3STCH_LyapVv_itop}\\
	V_{\tilde{P}\tilde{P}} & =4\left\Vert \tilde{P}\right\Vert ^{2}\mathbf{I}_{3}+8\tilde{P}\tilde{P}^{\top}\label{eq:SE3STCH_eq:LyapVvv_itop}
\end{align}
The first part of the differential operator $\mathcal{L}V$ in \eqref{eq:SE3STCH_LV_strat}
can be evaluated by 
\begin{align}
	V_{\tilde{\rho}}^{\top}\mathcal{F}_{\tilde{\rho}}= & -2\frac{\left\Vert \tilde{\rho}\right\Vert ^{2}}{\left(1+\left\Vert \tilde{\rho}\right\Vert ^{2}\right)^{2}}\tilde{\rho}^{\top}\hat{R}\left(\tilde{b}_{\Omega}+k_{w}W_{\Omega}+\hat{R}^{\top}{\rm diag}\left(\left[\frac{1}{2}\frac{\boldsymbol{\Upsilon}_{a}\left(\tilde{R}\right)}{1-\left\Vert \tilde{R}\right\Vert _{I}}\right]\right)\hat{\sigma}_{\Omega}\right)+\frac{\left\Vert \tilde{\rho}\right\Vert ^{2}}{\left(1+\left\Vert \tilde{\rho}\right\Vert ^{2}\right)^{2}}\tilde{\rho}^{\top}\mathcal{\bar{Q}}_{\Omega}^{2}\tilde{\rho}\nonumber \\
	\leq & -2\frac{\left\Vert \tilde{\rho}\right\Vert ^{2}}{\left(1+\left\Vert \tilde{\rho}\right\Vert ^{2}\right)^{2}}\tilde{\rho}^{\top}\hat{R}\left(\tilde{b}_{\Omega}+k_{w}W_{\Omega}-\hat{R}^{\top}{\rm diag}\left(\left[\frac{1}{2}\frac{\boldsymbol{\Upsilon}_{a}\left(\tilde{R}\right)}{1-\left\Vert \tilde{R}\right\Vert _{I}}\right]\right)\tilde{\sigma}_{\Omega}\right)\label{eq:SE3STCH_Vdot_strat}
\end{align}
Hence, the differential operator $\mathcal{L}V$ in \eqref{eq:SE3STCH_LV_strat}
can be described by 
\begin{align}
	\mathcal{L}V\leq & -4\tilde{X}^{\top}\left[\begin{array}{cc}
		\frac{\left\Vert \tilde{\rho}\right\Vert ^{2}}{\left(1+\left\Vert \tilde{\rho}\right\Vert ^{2}\right)^{3}}\mathbf{I}_{3} & \mathbf{0}_{3\times3}\\
		\mathbf{0}_{3\times3} & \left\Vert \tilde{P}\right\Vert ^{2}\mathbf{I}_{3}
	\end{array}\right]\mathcal{G}\left(\tilde{\rho}\right)\left(\breve{\overline{\mathbf{Ad}}}_{\hat{\boldsymbol{T}}}\left(\tilde{b}+k_{w}W\right)-\left[\begin{array}{cc}
		\frac{1}{2}\mathbf{I}_{3} & \mathbf{0}_{3\times3}\\
		\mathbf{0}_{3\times3} & \mathbf{0}_{3\times3}
	\end{array}\right]{\rm diag}\left(\tilde{X}\right)\tilde{\sigma}\right)\nonumber \\
	& +{\rm Tr}\left\{ \frac{\left\Vert \tilde{\rho}\right\Vert ^{4}\mathbf{I}_{3}+\left(\left\Vert \tilde{\rho}\right\Vert ^{2}\mathbf{I}_{3}+2\tilde{\rho}\tilde{\rho}^{\top}\right)}{2\left(1+\left\Vert \tilde{\rho}\right\Vert ^{2}\right)^{3}}\mathcal{\bar{Q}}_{\Omega}^{2}+2\left(\left\Vert \tilde{P}\right\Vert ^{2}\mathbf{I}_{3}+2\tilde{R}^{\top}\tilde{P}\tilde{P}^{\top}\tilde{R}\right)\mathcal{\bar{Q}}_{V}^{2}\right\} \nonumber \\
	& -\tilde{b}^{\top}\Gamma^{-1}\dot{\hat{b}}-\tilde{\sigma}^{\top}\Pi^{-1}\dot{\hat{\sigma}}-\frac{\left\Vert \tilde{\rho}\right\Vert ^{2}\left(1+3\left\Vert \tilde{\rho}\right\Vert ^{2}\right)\tilde{\rho}^{\top}\mathcal{\bar{Q}}_{\Omega}^{2}\tilde{\rho}}{2\left(1+\left\Vert \tilde{\rho}\right\Vert ^{2}\right)^{3}}\label{eq:SE3STCH_LV_strat1}
\end{align}
where $\frac{1}{4}\frac{\boldsymbol{\Upsilon}_{a}\left(\tilde{R}\right)}{1-\left\Vert \tilde{R}\right\Vert _{I}}=\frac{1}{2}\tilde{\rho}$
as given in \eqref{eq:SE3STCH_TR2} and \eqref{eq:SE3STCH_VEX_Pa}.
Now, let us simplify the trace bracket in \eqref{eq:SE3STCH_LV_strat1}.
To simplify the result in \eqref{eq:SE3STCH_LV_strat1}, one has 
\[
{\rm Tr}\left\{ \left(\left\Vert \tilde{\rho}\right\Vert ^{2}\mathbf{I}_{3}+2\tilde{\rho}\tilde{\rho}^{\top}\right)\mathcal{\bar{Q}}_{\Omega}^{2}\right\} \leq3\left\Vert \tilde{\rho}\right\Vert ^{2}{\rm Tr}\left\{ \mathcal{\bar{Q}}_{\Omega}^{2}\right\} 
\]
and for 
\[
\bar{q}_{\Omega}=\left[\mathcal{\bar{Q}}_{\Omega\left(1,1\right)},\mathcal{\bar{Q}}_{\Omega\left(2,2\right)},\bar{\mathcal{Q}}_{\Omega\left(3,3\right)}\right]^{\top}
\]
we have 
\[
\left\Vert \tilde{\rho}\right\Vert ^{2}{\rm Tr}\left\{ \mathcal{\bar{Q}}_{\Omega}^{2}\right\} =3\left\Vert \tilde{\rho}\right\Vert ^{2}\left\Vert \bar{q}_{\Omega}\right\Vert ^{2}
\]
Similarly, one can find 
\[
{\rm Tr}\left\{ \left(4\left\Vert \tilde{P}\right\Vert ^{2}\mathbf{I}_{3}+8\tilde{R}^{\top}\tilde{P}\tilde{P}^{\top}\tilde{R}\right)\mathcal{\bar{Q}}_{V}^{2}\right\} \leq12\left\Vert \tilde{P}\right\Vert ^{2}{\rm Tr}\left\{ \mathcal{\bar{Q}}_{V}^{2}\right\} 
\]
and for 
\[
\bar{q}_{V}=\left[\mathcal{\bar{Q}}_{V\left(1,1\right)},\mathcal{\bar{Q}}_{V\left(2,2\right)},\mathcal{\bar{Q}}_{V\left(3,3\right)}\right]^{\top}
\]
we have 
\[
12\left\Vert \tilde{P}\right\Vert ^{2}{\rm Tr}\left\{ \mathcal{\bar{Q}}_{V}^{2}\right\} =12\left\Vert \tilde{P}\right\Vert ^{2}\left\Vert \bar{q}_{V}\right\Vert ^{2}
\]
Hence, the operator in \eqref{eq:SE3STCH_LV_strat1} becomes 
\begin{align}
	\mathcal{L}V\leq & -4\tilde{X}^{\top}\left[\begin{array}{cc}
		\frac{\left\Vert \tilde{\rho}\right\Vert ^{2}}{\left(1+\left\Vert \tilde{\rho}\right\Vert ^{2}\right)^{3}}\mathbf{I}_{3} & \mathbf{0}_{3\times3}\\
		\mathbf{0}_{3\times3} & \left\Vert \tilde{P}\right\Vert ^{2}\mathbf{I}_{3}
	\end{array}\right]\mathcal{G}\left(\tilde{\rho}\right)\left(\breve{\overline{\mathbf{Ad}}}_{\hat{\boldsymbol{T}}}\left(\tilde{b}+k_{w}W\right)-\left[\begin{array}{cc}
		\frac{1}{2}\mathbf{I}_{3} & \mathbf{0}_{3\times3}\\
		\mathbf{0}_{3\times3} & \mathbf{0}_{3\times3}
	\end{array}\right]{\rm diag}\left(\tilde{X}\right)\tilde{\sigma}\right)\nonumber \\
	& +\frac{\left\Vert \tilde{\rho}\right\Vert ^{4}{\rm Tr}\left\{ \mathcal{\bar{Q}}_{\Omega}^{2}\right\} +3\left\Vert \tilde{\rho}\right\Vert ^{2}\left\Vert \bar{q}_{\Omega}\right\Vert ^{2}}{2\left(1+\left\Vert \tilde{\rho}\right\Vert ^{2}\right)^{3}}+6\left\Vert \tilde{P}\right\Vert ^{2}\left\Vert \bar{q}_{V}\right\Vert ^{2}-\tilde{b}^{\top}\Gamma^{-1}\dot{\hat{b}}-\tilde{\sigma}^{\top}\Pi^{-1}\dot{\hat{\sigma}}-\frac{\left\Vert \tilde{\rho}\right\Vert ^{2}\left(1+3\left\Vert \tilde{\rho}\right\Vert ^{2}\right)\tilde{\rho}^{\top}\mathcal{\bar{Q}}_{\Omega}^{2}\tilde{\rho}}{2\left(1+\left\Vert \tilde{\rho}\right\Vert ^{2}\right)^{3}}\label{eq:SE3STCH_LV_strat2}
\end{align}
According to Lemma \ref{lem:SE3STCH_2}, the following two equations
hold 
\begin{align}
	\frac{3\left\Vert \tilde{\rho}\right\Vert ^{2}\left\Vert \overline{q}_{\Omega}\right\Vert ^{2}}{2\left(1+\left\Vert \tilde{\rho}\right\Vert ^{2}\right)^{3}} & \leq\frac{1}{2\varepsilon}\frac{9}{4\left(1+\left\Vert \tilde{\rho}\right\Vert ^{2}\right)^{6}}\left\Vert \tilde{\rho}\right\Vert ^{4}+\frac{\varepsilon}{2}\left\Vert \bar{q}_{\Omega}\right\Vert ^{4}\nonumber \\
	& \leq\frac{9}{8\left(1+\left\Vert \tilde{\rho}\right\Vert ^{2}\right)^{3}\varepsilon}\left\Vert \tilde{\rho}\right\Vert ^{4}+\frac{\varepsilon}{2}\left(\sum_{i=1}^{3}\sigma_{i}\right)^{2}\label{eq:SE3STCH_Inq_Lyap}
\end{align}
\begin{align}
	6\left\Vert \tilde{P}\right\Vert ^{2}\left\Vert \overline{q}_{V}\right\Vert ^{2} & \leq\frac{36}{2\varepsilon}\left\Vert \tilde{P}\right\Vert ^{4}+\frac{\varepsilon}{2}\left\Vert \bar{q}_{V}\right\Vert ^{4}\nonumber \\
	& \leq\frac{18}{\varepsilon}\left\Vert \tilde{P}\right\Vert ^{4}+\frac{\varepsilon}{2}\left(\sum_{i=4}^{6}\sigma_{i}\right)^{2}\label{eq:SE3STCH_Inq_Lyap-1}
\end{align}
Considering the results in \eqref{eq:SE3STCH_Inq_Lyap} and \eqref{eq:SE3STCH_Inq_Lyap-1},
in addition, $\left(\sum_{i=1}^{6}\sigma_{i}\right)^{2}\geq\left(\sum_{i=1}^{3}\sigma_{i}\right)^{2}+\left(\sum_{i=4}^{6}\sigma_{i}\right)^{2}$,
hence, the operator in \eqref{eq:SE3STCH_LV_strat2} can be expressed
as 
\begin{align}
	\mathcal{L}V\leq & -4\tilde{X}^{\top}\left[\begin{array}{cc}
		\frac{\left\Vert \tilde{\rho}\right\Vert ^{2}}{\left(1+\left\Vert \tilde{\rho}\right\Vert ^{2}\right)^{3}}\mathbf{I}_{3} & \mathbf{0}_{3\times3}\\
		\mathbf{0}_{3\times3} & \left\Vert \tilde{P}\right\Vert ^{2}\mathbf{I}_{3}
	\end{array}\right]\mathcal{G}\left(\tilde{\rho}\right)\left(\breve{\overline{\mathbf{Ad}}}_{\hat{\boldsymbol{T}}}\left(\tilde{b}+k_{w}W\right)-\left[\begin{array}{cc}
		\frac{1}{2}\mathbf{I}_{3} & \mathbf{0}_{3\times3}\\
		\mathbf{0}_{3\times3} & \mathbf{0}_{3\times3}
	\end{array}\right]{\rm diag}\left(\tilde{X}\right)\tilde{\sigma}\right)\nonumber \\
	& +\frac{\left\Vert \tilde{\rho}\right\Vert ^{4}{\rm Tr}\left\{ \mathcal{\bar{Q}}_{\Omega}^{2}\right\} }{2\left(1+\left\Vert \tilde{\rho}\right\Vert ^{2}\right)^{3}}+\frac{9\left\Vert \tilde{\rho}\right\Vert ^{4}}{8\left(1+\left\Vert \tilde{\rho}\right\Vert ^{2}\right)^{3}\varepsilon}+\frac{18}{\varepsilon}\left\Vert \tilde{P}\right\Vert ^{4}+\frac{\varepsilon}{2}\left(\sum_{i=1}^{6}\sigma_{i}\right)^{2}-\tilde{b}^{\top}\Gamma^{-1}\dot{\hat{b}}-\tilde{\sigma}^{\top}\Pi^{-1}\dot{\hat{\sigma}}\nonumber \\
	& -\frac{\left\Vert \tilde{\rho}\right\Vert ^{2}\left(1+3\left\Vert \tilde{\rho}\right\Vert ^{2}\right)\tilde{\rho}^{\top}\mathcal{\bar{Q}}_{\Omega}^{2}\tilde{\rho}}{2\left(1+\left\Vert \tilde{\rho}\right\Vert ^{2}\right)^{3}}\label{eq:SE3STCH_LV_strat3}
\end{align}
The result in \eqref{eq:SE3STCH_LV_strat3} can be written as 
\begin{align}
	\mathcal{L}V\leq & -4\tilde{X}^{\top}\left[\begin{array}{cc}
		\frac{\left\Vert \tilde{\rho}\right\Vert ^{2}}{\left(1+\left\Vert \tilde{\rho}\right\Vert ^{2}\right)^{3}}\mathbf{I}_{3} & \mathbf{0}_{3\times3}\\
		\mathbf{0}_{3\times3} & \left\Vert \tilde{P}\right\Vert ^{2}\mathbf{I}_{3}
	\end{array}\right]\mathcal{G}\left(\tilde{\rho}\right)\left(\breve{\overline{\mathbf{Ad}}}_{\hat{\boldsymbol{T}}}\left(\tilde{b}+k_{w}W\right)-\left[\begin{array}{cc}
		\frac{1}{2}\mathbf{I}_{3} & \mathbf{0}_{3\times3}\\
		\mathbf{0}_{3\times3} & \mathbf{0}_{3\times3}
	\end{array}\right]{\rm diag}\left(\tilde{X}\right)\tilde{\sigma}\right)\nonumber \\
	& +\tilde{X}^{\top}\left[\begin{array}{cc}
		2\frac{\left\Vert \tilde{\rho}\right\Vert ^{2}}{\left(1+\left\Vert \tilde{\rho}\right\Vert ^{2}\right)^{2}}\mathbf{I}_{3} & \mathbf{0}_{3\times3}\\
		\mathbf{0}_{3\times3} & 4\left\Vert \tilde{P}\right\Vert ^{2}\mathbf{I}_{3}
	\end{array}\right]\left(\left[\begin{array}{cc}
		\frac{1}{4}\frac{\mathcal{D}_{\tilde{\rho}}}{1+\left\Vert \tilde{\rho}\right\Vert ^{2}} & \mathbf{0}_{3\times3}\\
		\mathbf{0}_{3\times3} & \mathbf{0}_{3\times3}
	\end{array}\right]\sigma+\frac{1}{\varepsilon}\left[\begin{array}{cc}
		\frac{9}{16}\frac{1}{1+\left\Vert \tilde{\rho}\right\Vert ^{2}}\mathbf{I}_{3} & \mathbf{0}_{3\times3}\\
		\mathbf{0}_{3\times3} & 4.5\mathbf{I}_{3}
	\end{array}\right]\tilde{X}\right)\nonumber \\
	& +\frac{\varepsilon}{2}\left(\sum_{i=1}^{6}\sigma_{i}\right)^{2}-\tilde{b}^{\top}\Gamma^{-1}\dot{\hat{b}}-\tilde{\sigma}^{\top}\Pi^{-1}\dot{\hat{\sigma}}-\frac{\left\Vert \tilde{\rho}\right\Vert ^{2}\left(1+3\left\Vert \tilde{\rho}\right\Vert ^{2}\right)\tilde{\rho}^{\top}\mathcal{\bar{Q}}_{\Omega}^{2}\tilde{\rho}}{2\left(1+\left\Vert \tilde{\rho}\right\Vert ^{2}\right)^{3}}\label{eq:SE3STCH_LV_strat4}
\end{align}
According to \eqref{eq:SE3STCH_TR2} and \eqref{eq:SE3STCH_VEX_Pa},
we have $\left\Vert \tilde{R}\right\Vert _{I}=\left\Vert \tilde{\rho}\right\Vert ^{2}/\left(1+\left\Vert \tilde{\rho}\right\Vert ^{2}\right)$
and $\boldsymbol{\Upsilon}_{a}\left(\tilde{R}\right)=2\tilde{\rho}/\left(1+\left\Vert \tilde{\rho}\right\Vert ^{2}\right)$,
while $\left\Vert \boldsymbol{\Upsilon}_{a}\left(\tilde{R}\right)\right\Vert ^{2}=4\left(1-\left\Vert \tilde{R}\right\Vert _{I}\right)\left\Vert \tilde{R}\right\Vert _{I}=4\frac{\left\Vert \tilde{\rho}\right\Vert ^{2}}{\left(1+\left\Vert \tilde{\rho}\right\Vert ^{2}\right)^{2}}$
as in \eqref{eq:SE3STCH_VEX2_Pa}. Substituting for the differential
operators $\dot{\hat{b}}$ and $\dot{\hat{\sigma}}$ and the correction
factor $W$ from \eqref{eq:SE3STCH_best_Strat}, \eqref{eq:SE3STCH_gest_strat}
and \eqref{eq:SE3STCH_Wcorr_Strat}, respectively, yields 
\begin{align}
	\mathcal{L}V\leq & -4\left(\left(k_{p}k_{w}-\frac{1}{8}\right)\left(\sum_{i=1}^{3}\sigma_{i}\right)+\frac{1}{\varepsilon}\left(k_{p}k_{w}-\frac{9}{32}\right)\right)\frac{\left\Vert \tilde{\rho}\right\Vert ^{4}}{\left(1+\left\Vert \tilde{\rho}\right\Vert ^{2}\right)^{3}}-\frac{\left\Vert \tilde{\rho}\right\Vert ^{2}\left(1+3\left\Vert \tilde{\rho}\right\Vert ^{2}\right)\tilde{\rho}^{\top}\mathcal{\bar{Q}}_{\Omega}^{2}\tilde{\rho}}{2\left(1+\left\Vert \tilde{\rho}\right\Vert ^{2}\right)^{3}}\nonumber \\
	& -\frac{4k_{p}k_{w}}{\varepsilon}\left(\frac{\left\Vert \tilde{\rho}\right\Vert ^{2}}{1+\left\Vert \tilde{\rho}\right\Vert ^{2}}\right)^{2}-4\left(k_{p}k_{w}-4.5\right)\left\Vert \tilde{P}\right\Vert ^{4}-k_{b}\left\Vert \tilde{b}\right\Vert ^{2}-k_{\sigma}\left\Vert \tilde{\sigma}\right\Vert ^{2}+k_{b}\tilde{b}^{\top}b+k_{\sigma}\tilde{\sigma}^{\top}\sigma+\frac{\varepsilon}{2}\left(\sum_{i=1}^{6}\sigma_{i}\right)^{2}\label{eq:SE3STCH_LV4_strat}
\end{align}
applying Young's inequality to $k_{b}\tilde{b}^{\top}b$ and $k_{\sigma}\tilde{\sigma}^{\top}\sigma$,
respectively, one has 
\begin{align*}
	k_{b}\tilde{b}^{\top}b & \leq\frac{k_{b}}{2}\left\Vert \tilde{b}\right\Vert ^{2}+\frac{k_{b}}{2}\left\Vert b\right\Vert ^{2}\\
	k_{\sigma}\tilde{\sigma}^{\top}\sigma & \leq\frac{k_{\sigma}}{2}\left\Vert \tilde{\sigma}\right\Vert ^{2}+\frac{k_{\sigma}}{2}\left(\sum_{i=1}^{6}\sigma_{i}\right)^{2}
\end{align*}
consequently, \eqref{eq:SE3STCH_LV4_strat} becomes 
\begin{align}
	\mathcal{L}V\leq & -4\left(\left(k_{p}k_{w}-\frac{1}{8}\right)\left(\sum_{i=1}^{3}\sigma_{i}\right)+\frac{1}{\varepsilon}\left(k_{p}k_{w}-\frac{9}{32}\right)\right)\frac{\left\Vert \tilde{\rho}\right\Vert ^{4}}{\left(1+\left\Vert \tilde{\rho}\right\Vert ^{2}\right)^{3}}-\frac{\left\Vert \tilde{\rho}\right\Vert ^{2}\left(1+3\left\Vert \tilde{\rho}\right\Vert ^{2}\right)\tilde{\rho}^{\top}\mathcal{\bar{Q}}_{\Omega}^{2}\tilde{\rho}}{2\left(1+\left\Vert \tilde{\rho}\right\Vert ^{2}\right)^{3}}\nonumber \\
	& -\frac{4k_{p}k_{w}}{\varepsilon}\left(\frac{\left\Vert \tilde{\rho}\right\Vert ^{2}}{1+\left\Vert \tilde{\rho}\right\Vert ^{2}}\right)^{2}-4\left(k_{p}k_{w}-4.5\right)\left\Vert \tilde{P}\right\Vert ^{4}-\frac{k_{b}}{2}\left\Vert \tilde{b}\right\Vert ^{2}-\frac{k_{\sigma}}{2}\left\Vert \tilde{\sigma}\right\Vert ^{2}+\frac{k_{b}}{2}\left\Vert b\right\Vert ^{2}+\frac{1}{2}\left(k_{\sigma}+\varepsilon\right)\left(\sum_{i=1}^{6}\sigma_{i}\right)^{2}\label{eq:SE3STCH_LV_Final_strat}
\end{align}
Setting $\gamma>0$, $\bar{\pi}>0$, $k_{p}k_{w}>4.5$, $k_{b}>0$,
$k_{\sigma}>0$, and the positive constant $\varepsilon$ is sufficiently
small, the operator $\mathcal{L}V$ in \eqref{eq:SE3STCH_LV4_strat}
becomes similar to (54) and (75) in \cite{hashim2018Stoch} or (4.16) in \cite{deng2001stabilization} which is
in turn similar to \eqref{eq:SE3STCH_dVfunction_Lyap} in Lemma \ref{lem:SE3STCH_1}.
In that case, the constant component $c_{2}$ in Lemma \ref{lem:SE3STCH_1}
is $c_{2}=\frac{k_{b}}{2}\left\Vert b\right\Vert ^{2}+\frac{1}{2}\left(k_{\sigma}+\varepsilon\right)\left(\sum_{i=1}^{6}\sigma_{i}\right)^{2}$.
Let us define 
\begin{align*}
	c_{2}= & \frac{k_{b}}{2}\left\Vert b\right\Vert ^{2}+\frac{1}{2}\left(k_{\sigma}+\varepsilon\right)\left(\sum_{i=1}^{6}\sigma_{i}\right)^{2}\\
	\tilde{Y}= & \left[\frac{\left\Vert \tilde{\rho}\right\Vert ^{2}}{1+\left\Vert \tilde{\rho}\right\Vert ^{2}},\left\Vert \tilde{P}\right\Vert ^{2},\frac{1}{\sqrt{2\gamma}}\tilde{b}^{\top},\frac{1}{\sqrt{2\bar{\pi}}}\tilde{\sigma}^{\top}\right]^{\top}\in\mathbb{R}^{14},\\
	\mathcal{H}= & {\rm diag}\left(\frac{4k_{p}k_{w}}{\varepsilon},4\left(k_{p}k_{w}-4.5\right),\gamma k_{b}\underline{\mathbf{1}}_{6}^{\top},\bar{\pi}k_{\sigma}\underline{\mathbf{1}}_{6}^{\top}\right)\in\mathbb{R}^{14\times14}
\end{align*}
The differential operator in \eqref{eq:SE3STCH_LV_Final_strat} is
\begin{align}
	\mathcal{L}V\leq & -4\left(\left(k_{p}k_{w}-\frac{1}{8}\right)\left(\sum_{i=1}^{3}\sigma_{i}\right)+\frac{1}{\varepsilon}\left(k_{p}k_{w}-\frac{9}{32}\right)\right)\frac{\left\Vert \tilde{\rho}\right\Vert ^{4}}{\left(1+\left\Vert \tilde{\rho}\right\Vert ^{2}\right)^{3}}-\frac{\left\Vert \tilde{\rho}\right\Vert ^{2}\left(1+3\left\Vert \tilde{\rho}\right\Vert ^{2}\right)\tilde{\rho}^{\top}\mathcal{\bar{Q}}_{\Omega}^{2}\tilde{\rho}}{2\left(1+\left\Vert \tilde{\rho}\right\Vert ^{2}\right)^{3}}\nonumber \\
	& -\tilde{Y}^{\top}\mathcal{H}\tilde{Y}+c_{2}\label{eq:SE3STCH_LV5_strat}
\end{align}
and more simply 
\begin{equation}
	\mathcal{L}V\leq-h\left(\left\Vert \tilde{\rho}\right\Vert \right)-\underline{\lambda}\left(\mathcal{H}\right)V+c_{2}\label{eq:SE3STCH_LV6_strat}
\end{equation}
such that $h\left(\cdot\right)$ is a class $\mathcal{K}$ function
that includes the first two components in \eqref{eq:SE3STCH_LV5_strat},
and $\underline{\lambda}\left(\cdot\right)$ denotes the minimum eigenvalue
of a matrix. Based on \eqref{eq:SE3STCH_LV6_strat}, one easily obtains
\begin{equation}
	\frac{d\left(\mathbb{E}\left[V\right]\right)}{dt}=\mathbb{E}\left[\mathcal{L}V\right]\leq-\underline{\lambda}\left(\mathcal{H}\right)V+c_{2}\label{eq:SE3STCH_dV_Exp_strat}
\end{equation}
as such, \eqref{eq:SE3STCH_dV_Exp_strat} means that 
\begin{equation}
	0\leq\mathbb{E}\left[V\left(t\right)\right]\leq V\left(0\right){\rm exp}\left(-\underline{\lambda}\left(\mathcal{H}\right)t\right)+\frac{c_{2}}{\underline{\lambda}\left(\mathcal{H}\right)},\,\forall t\geq0\label{eq:SE3STCH_V_Exp_strat}
\end{equation}
The inequality in \eqref{eq:SE3STCH_V_Exp_strat} implies that $\mathbb{E}\left[V\left(t\right)\right]$
is eventually bounded by $c_{2}/\underline{\lambda}\left(\mathcal{H}\right)$.
Since, $\mathcal{Q}^{2}:\mathbb{R}_{+}\rightarrow\mathbb{R}^{6\times6}$
is bounded, the operator in \eqref{eq:SE3STCH_dV_Exp_strat} is $\mathcal{L}V\leq c_{2}/\underline{\lambda}\left(\mathcal{H}\right)$.
Define $\tilde{Z}=\left[\tilde{\rho}^{\top},\tilde{P}^{\top},\tilde{b}^{\top},\tilde{\sigma}^{\top}\right]^{\top}\in\mathbb{R}^{18}$,
$\tilde{Z}$ is SGUUB in mean square as in Definition \ref{def:SE3STCH_1}.
Define $\mathcal{U}_{0}\subseteq\mathbb{SO}\left(3\right)\times\mathbb{R}^{3}\times\mathbb{R}^{6}\times\mathbb{R}^{6}$
by 
\[
\mathcal{U}_{0}=\left\{ \left.\left(\tilde{R}\left(0\right),\tilde{P}\left(0\right),\tilde{b}\left(0\right),\tilde{\sigma}\left(0\right)\right)\right|{\rm Tr}\left\{ \tilde{R}\left(0\right)\right\} =-1,\tilde{P}\left(0\right)=\underline{\mathbf{0}}_{3},\tilde{b}\left(0\right)=\underline{\mathbf{0}}_{6},\tilde{\sigma}\left(0\right)=\underline{\mathbf{0}}_{6}\right\} 
\]
The set $\mathcal{U}_{0}$ is forward invariant and unstable. Therefore,
from almost any initial condition such that $\tilde{R}\left(0\right)\notin\mathcal{U}_{0}$
or equivalently for any $\tilde{\rho}\left(0\right)\in\mathbb{R}^{3}$,
the trajectory of $\tilde{Z}$ converges to the neighborhood of the
origin which depends on the value of $c_{2}/\underline{\lambda}\left(\mathcal{H}\right)$
in \eqref{eq:SE3STCH_V_Exp_strat}. %
{} From Lemma \ref{lem:SE3STCH_1} and design parameters of the stochastic
observer in Theorem \ref{thm:SE3STCH_2} in addition if we have prior
knowledge about the covariance upper bound $\sigma$, $c_{2}/\underline{\lambda}\left(\mathcal{H}\right)$
can be made smaller if we choose the design parameters appropriately.
Clearly, the minimum singular value of $\underline{\lambda}\left(\mathcal{H}\right)$
can be controlled by $k_{p}$, $k_{w}$, $\gamma$ and $\bar{\pi}$.
To conclude our discussion, it should be remarked that solving the
problem in the sense of Stratonovich with the proper selection of
potential function as in \eqref{eq:SE3STCH_LyapV_strat} helps to
attenuate or control the noise level associated with the velocity
measurements vector $\mathcal{Y}_{m}$. The proposed nonlinear stochastic
filter is able to correct the position as well as the attitude and
reduce the noise level associated with velocity measurements $\mathcal{Y}_{m}$
through the setting of parameters in presence of high level of noise
and bias components. This advantage is not given in nonlinear deterministic
$\mathbb{SE}\left(3\right)$ filters. The main benefit of the nonlinear
stochastic filter in the sense of Stratonovich is that no prior information
about the covariance matrix $\mathcal{Q}^{2}$ is required. Also,
the filter is applicable for white as well as colored noise which
offers flexibility in the design process. 
\begin{rem}
	Notice that, as $k_{p},k_{w},\gamma,\bar{\pi}\rightarrow\infty$ and
	$\varepsilon\rightarrow0$, $\mathbb{P}\left\{ \underset{t\rightarrow\infty}{{\rm lim}}\left\Vert \tilde{X}\right\Vert =0\right\} \rightarrow1,\forall t\geq0$
	with perfect cancellation of undesirable time-variant components and
	uncertainties. 
\end{rem}

\section{Simulations \label{sec:SE3STCH_Simulations}}

This section presents the performance of the proposed nonlinear stochastic
filter on $\mathbb{SE}\left(3\right)$ considering high levels of
bias and noise introduced in the measurement process combined with
the large initial error in the homogeneous transformation matrix $\tilde{\boldsymbol{T}}\left(0\right)$.
The performance of the proposed stochastic filter is compared to \cite{hua2011observer}.
Let us define the dynamics of the homogeneous transformation matrix
$\boldsymbol{T}$ as in \eqref{eq:SE3STCH_T_Dynamics}. Let the angular
velocity input signal be 
\[
\Omega=\left[\begin{array}{c}
{\rm sin}\left(0.3t\right)\\
0.7{\rm sin}\left(0.25t+\pi\right)\\
0.5{\rm sin}\left(0.2t+\frac{\pi}{3}\right)
\end{array}\right]\left({\rm rad/sec}\right)
\]
with initial attitude being $R\left(0\right)=\mathbf{I}_{3}$. Let
the translational velocity be 
\[
V=\left[\begin{array}{c}
{\rm sin}\left(0.2t\right)\\
0.6{\rm sin}\left(0.15t+\frac{\pi}{2}\right)\\
{\rm sin}\left(0.25t+\frac{\pi}{4}\right)
\end{array}\right]\left({\rm m/sec}\right)
\]
and the initial position $P\left(0\right)=\underline{\mathbf{0}}_{3}$.
Let the angular velocity measurement $\Omega_{m}=\Omega+b_{\Omega}+\omega_{\Omega}$
be corrupted with a wide-band of random noise process with zero mean
$\omega_{\Omega}$ and standard deviation (STD) equal to $0.15\left({\rm rad/sec}\right)$
and $b_{\Omega}=0.1\left[1,-1,1\right]^{\top}$. Similarly, let the
translational velocity measurement $V_{m}=V+b_{V}+\omega_{V}$ be
subject to a wide-band of random noise process $\omega_{V}$ with
zero mean and ${\rm STD}=0.15\left({\rm m/sec}\right)$, and $b_{V}=0.1\left[2,5,1\right]^{\top}$.

Consider one landmark feature available for measurement $\left(N_{{\rm L}}=1\right)$
\[
{\rm v}_{1}^{\mathcal{I}\left({\rm L}\right)}=\left[\frac{1}{2},\sqrt{2},1\right]^{\top}
\]
and body-frame measurements obtained by \eqref{eq:SE3STCH_Vec_Landmark}
such that 
\[
{\rm v}_{i}^{\mathcal{B}\left({\rm L}\right)}=R^{\top}\left({\rm v}_{i}^{\mathcal{I}\left({\rm L}\right)}-P\right)+{\rm b}_{i}^{\mathcal{B}\left({\rm L}\right)}+\omega_{i}^{\mathcal{B}\left({\rm L}\right)}
\]
where the bias vector is defined as ${\rm b}_{1}^{\mathcal{B}\left({\rm L}\right)}=0.1\left[1.5,1,-1\right]^{\top}$
and a Gaussian noise vector $\omega_{1}^{\mathcal{B}\left({\rm L}\right)}$
with zero mean and ${\rm STD}=0.1$ corrupts the body-frame vector
measurements associated with the feature point.

Consider that two non-collinear inertial-frame vectors $\left(N_{{\rm R}}=2\right)$
are given by 
\begin{align*}
	{\rm v}_{1}^{\mathcal{I}\left({\rm R}\right)} & =\frac{1}{\sqrt{3}}\left[1,-1,1\right]^{\top}\\
	{\rm v}_{2}^{\mathcal{I}\left({\rm R}\right)} & =\left[0,0,1\right]^{\top}
\end{align*}
while body-frame vectors ${\rm v}_{1}^{\mathcal{B}\left({\rm R}\right)}$
and ${\rm v}_{2}^{\mathcal{B}\left({\rm R}\right)}$ are obtained
by \eqref{eq:SE3STCH_Vect_R} 
\[
{\rm v}_{i}^{\mathcal{B}\left({\rm R}\right)}=R^{\top}{\rm v}_{i}^{\mathcal{I}\left({\rm R}\right)}+{\rm b}_{i}^{\mathcal{B}\left({\rm R}\right)}+\omega_{i}^{\mathcal{B}\left({\rm R}\right)}
\]
for $i=1,2$. The body-frame vector measurements are subject to bias
components ${\rm b}_{1}^{\mathcal{B}\left({\rm R}\right)}=0.1\left[-1,1,0.5\right]^{\top}$
and ${\rm b}_{2}^{\mathcal{B}\left({\rm R}\right)}=0.1\left[0,0,1\right]^{\top}$.
In addition to bias, Gaussian noise vectors $\omega_{1}^{\mathcal{B}\left({\rm R}\right)}$
and $\omega_{2}^{\mathcal{B}\left({\rm R}\right)}$ with zero mean
and of ${\rm STD}=0.1$ corrupt the measurements. The third inertial
and body-frame vector measurements are obtained by ${\rm v}_{3}^{\mathcal{I}\left({\rm R}\right)}={\rm v}_{1}^{\mathcal{I}\left({\rm R}\right)}\times{\rm v}_{2}^{\mathcal{I}\left({\rm R}\right)}$
and ${\rm v}_{3}^{\mathcal{B}\left({\rm R}\right)}={\rm v}_{1}^{\mathcal{B}\left({\rm R}\right)}\times{\rm v}_{2}^{\mathcal{B}\left({\rm R}\right)}$.
Next, both body-frame and inertial-frame vectors are normalized, such
that ${\rm v}_{i}^{\mathcal{B}\left({\rm R}\right)}$ and ${\rm v}_{i}^{\mathcal{I}\left({\rm R}\right)}$
are normalized to $\upsilon_{i}^{\mathcal{B}\left({\rm R}\right)}$
and $\upsilon_{i}^{\mathcal{I}\left({\rm R}\right)}$, respectively,
for $i=1,2,3$ as given in \eqref{eq:SE3STCH_Vector_norm}. Therefore,
Assumption \ref{Assum:SE3STCH_1} holds. From vectorial measurements,
the corrupted reconstructed attitude $R_{y}$ is obtained by SVD \cite{markley1988attitude}
with $\tilde{R}=R_{y}\hat{R}^{\top}$, \nameref{sec:SE3STCH_AppendixB}.
The total simulation time is 30 seconds.

For large initial attitude error, the initial rotation of attitude
estimate is given according to the mapping of angle-axis parameterization
in \eqref{eq:SE3STCH_att_ang} by $\hat{R}\left(0\right)=\mathcal{R}_{\alpha}\left(\alpha,u/\left\Vert u\right\Vert \right)$
with $\alpha=170\left({\rm deg}\right)$ and $u=\left[3,10,8\right]^{\top}$
such that $\left\Vert \tilde{R}\left(0\right)\right\Vert _{I}$ approaches
the unstable equilibria $+1$. Also, the initial position of the estimator
is selected to be $\hat{P}\left(0\right)=\left[2,3,1\right]^{\top}$.
The matrices below summarize the initial conditions: 
\[
\boldsymbol{T}\left(0\right)=\left[\begin{array}{cccc}
1 & 0 & 0 & 0\\
0 & 1 & 0 & 0\\
0 & 0 & 1 & 0\\
0 & 0 & 0 & 1
\end{array}\right],\hspace{1em}\hat{\boldsymbol{T}}\left(0\right)=\left[\begin{array}{cccc}
-0.8816 & 0.2386 & 0.4074 & 2\\
0.4498 & 0.1625 & 0.8782 & 3\\
0.1433 & 0.9574 & -0.2505 & 1\\
0 & 0 & 0 & 1
\end{array}\right]
\]

The initial estimates of $\hat{b}$ and $\hat{\sigma}$ are $\hat{b}\left(0\right)=\underline{\mathbf{0}}_{6}$
and $\hat{\sigma}\left(0\right)=\underline{\mathbf{0}}_{6}$. Design
parameters used in the derivation of the nonlinear stochastic filter
are selected as $\Gamma=\mathbf{I}_{6}$, $\Pi=\mathbf{I}_{6}$, $k_{b}=0.1$,
$k_{\sigma}=0.1$, $k_{p}=2$, $k_{w}=3$, and $\varepsilon=0.5$.
Additionally, the following color notation is used: green color refers
to the true value, blue represents the performance of the proposed
nonlinear stochastic filter, and red illustrates the performance of
the filter previously proposed in literature. Finally, magenta demonstrates
measured values.

\begin{figure}[h]
	\centering{}\includegraphics[scale=0.35]{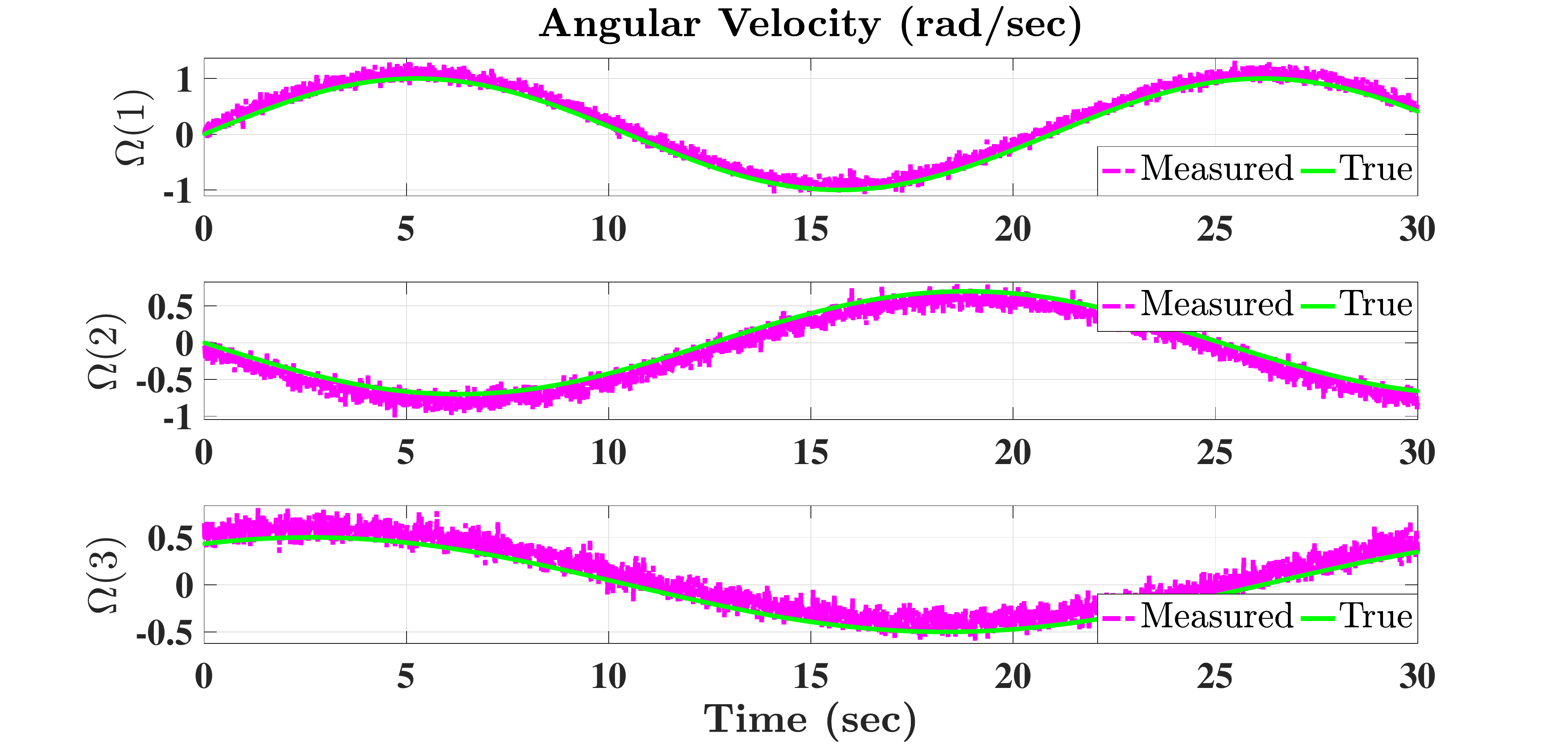}\caption{True and measured angular velocities.}
	\label{fig:SE3STCH_2} 
\end{figure}

\begin{figure}[h]
	\centering{}\includegraphics[scale=0.35]{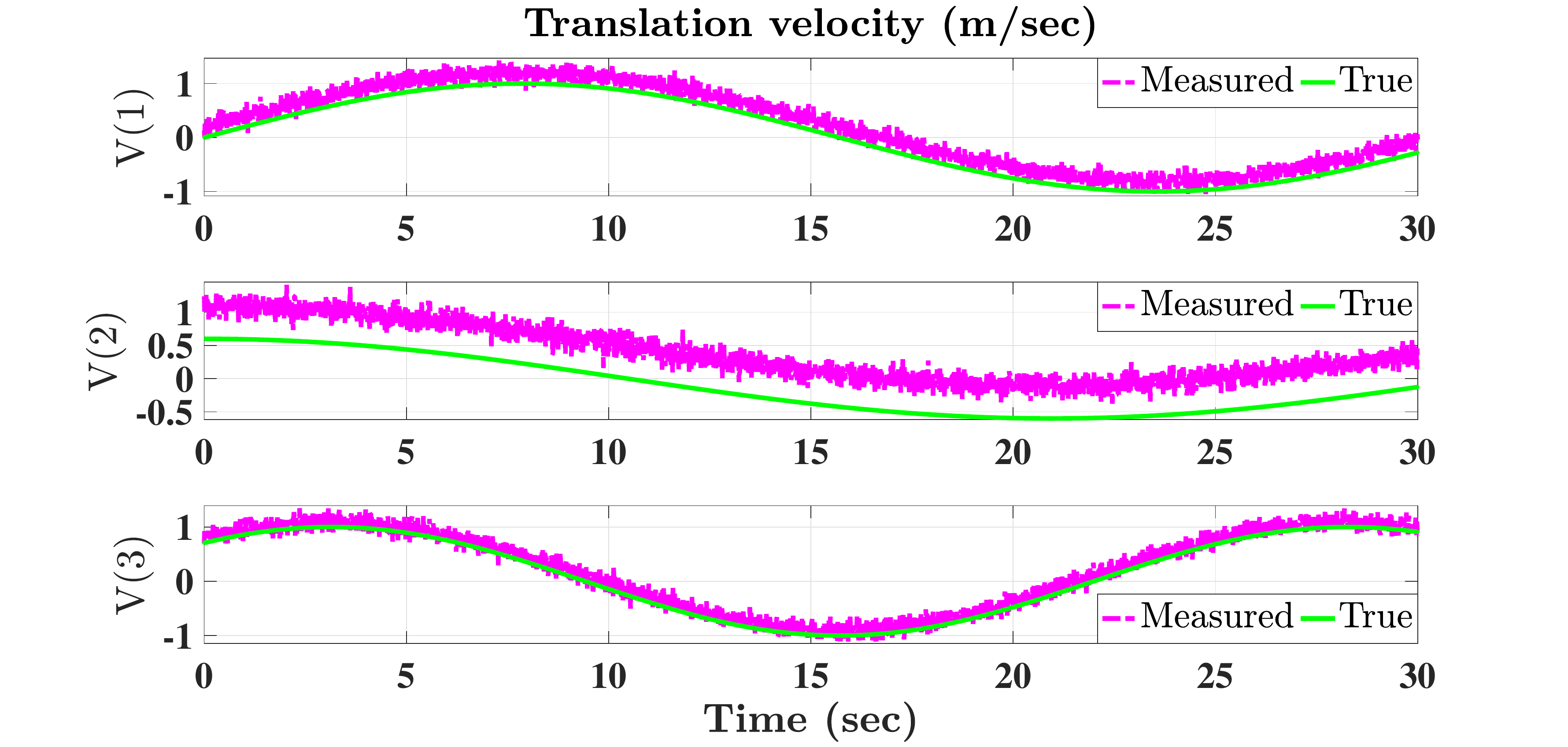}\caption{True and measured translational velocities.}
	\label{fig:SE3STCH_3} 
\end{figure}

\begin{figure}[h]
	\centering{}\includegraphics[scale=0.35]{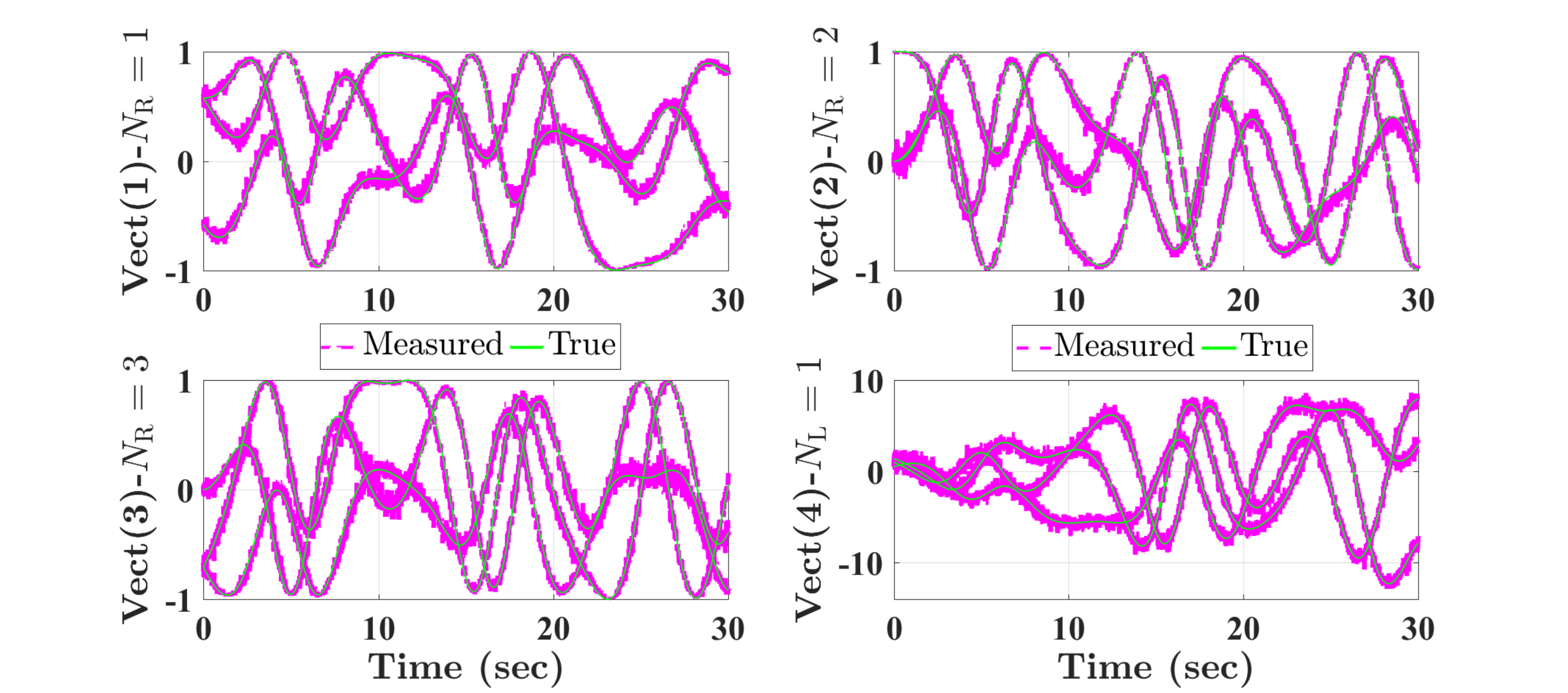}\caption{True values and vectorial measurements of the body-frame.}
	\label{fig:SE3STCH_4} 
\end{figure}

The first three figures present the true values of the velocity vectors
and body-frame vectors plotted against their measured values. The
true angular velocity $\left(\Omega\right)$ and the high values of
noise and bias components introduced through the measurement process
of $\Omega_{m}$ plotted against time are depicted in Figure \ref{fig:SE3STCH_2}.
Similarly, the true translational velocity $\left(V\right)$ and the
high values of noise and bias components associated with the measurement
process of $V_{m}$ plotted against time are illustrated in Figure
\ref{fig:SE3STCH_3}. In addition, Figure \ref{fig:SE3STCH_4} presents
the true body-frame vectors and their uncertain measurements corrupted
with noise. High levels of noise and bias inherent to the measurements
can be noticed in all the above-mentioned graphs (Figure \ref{fig:SE3STCH_2},
\ref{fig:SE3STCH_3} and \ref{fig:SE3STCH_4}). 

\begin{figure}[h]
	\centering{}\includegraphics[scale=0.35]{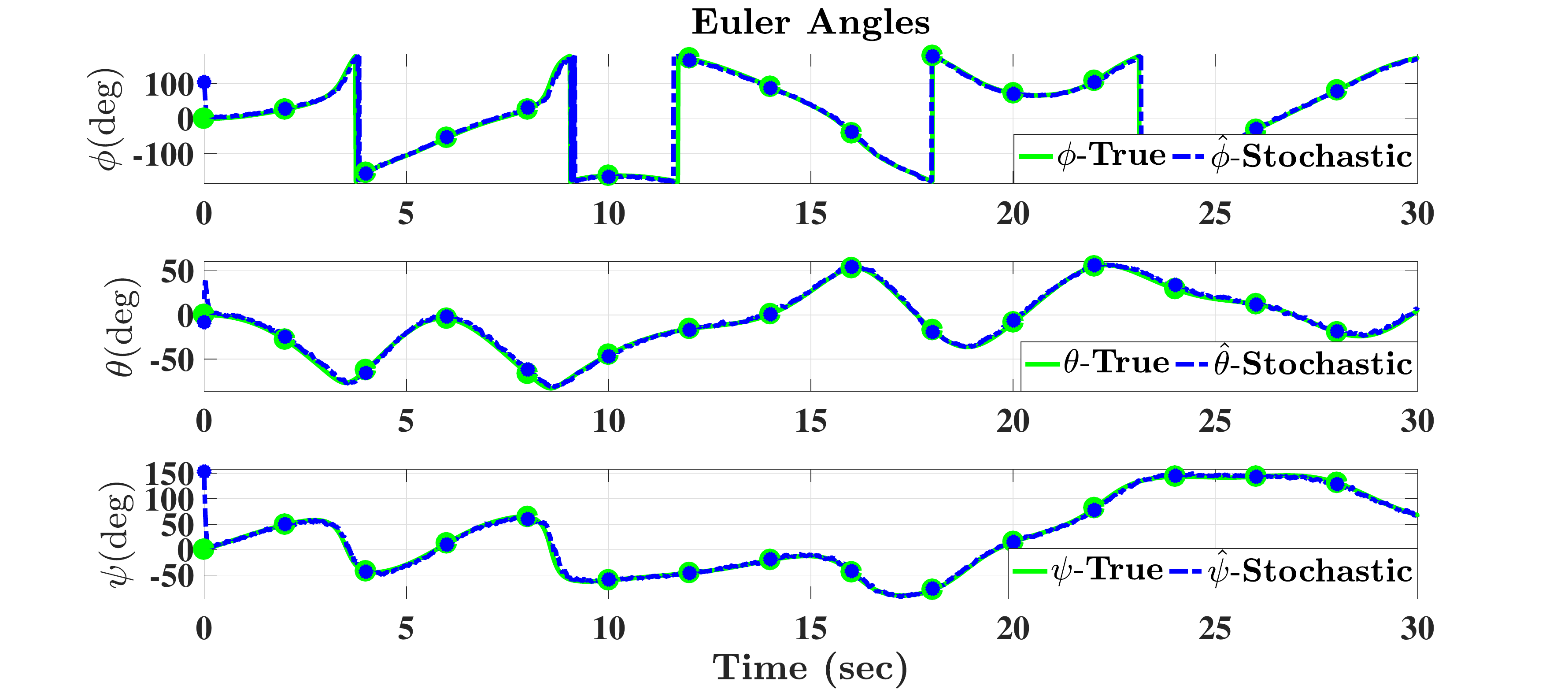}\caption{Tracking performance of Euler angles of the stochastic filter.}
	\label{fig:SE3STCH_5} 
\end{figure}

\begin{figure}[h]
	\centering{}\includegraphics[scale=0.35]{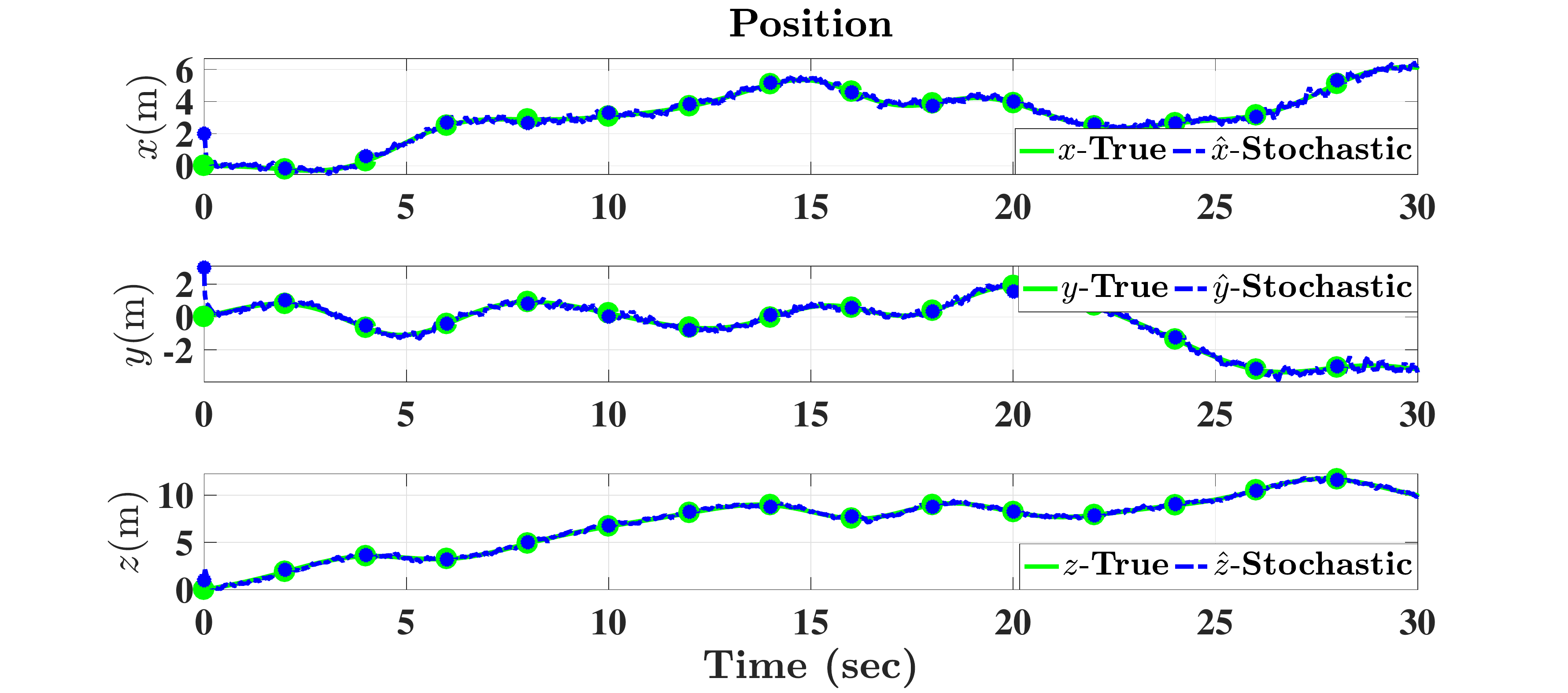}\caption{Tracking performance of $x$, $y$ and $z$ trajectory of the stochastic
		filter in 3D space.}
	\label{fig:SE3STCH_6} 
\end{figure}

\newpage{}The position and attitude tracking performance of the proposed
stochastic filter is demonstrated in Figure \ref{fig:SE3STCH_5} and
\ref{fig:SE3STCH_6}. Figure \ref{fig:SE3STCH_5} depicts the estimated
Euler angles $\left({\rm \text{Roll}}\left(\hat{\phi}\right),{\rm \text{Pitch}}\left(\hat{\theta}\right),{\rm \text{Yaw}}\left(\hat{\psi}\right)\right)$
versus the true values $\left(\phi,\theta,\psi\right)$. Also, Figure
\ref{fig:SE3STCH_6} illustrates the high value of the attitude initial
error. The tracking position $\left(\hat{x},\hat{y},\hat{z}\right)$
of the stochastic estimator in 3D space is compared to the true position
$\left(x,y,z\right)$ over time in Figure \ref{fig:SE3STCH_6}. Figure
\ref{fig:SE3STCH_5} and \ref{fig:SE3STCH_6} show impressive tracking
performance of the proposed stochastic observer in terms of position
and attitude in presence of large initial error between the true and
the estimated pose. Also, Figure \ref{fig:SE3STCH_5} and \ref{fig:SE3STCH_6}
demonstrate remarkable tracking performance in case when high values
of bias and noise corrupt the measurements. 

\begin{figure}[h]
	\centering{}\includegraphics[scale=0.35]{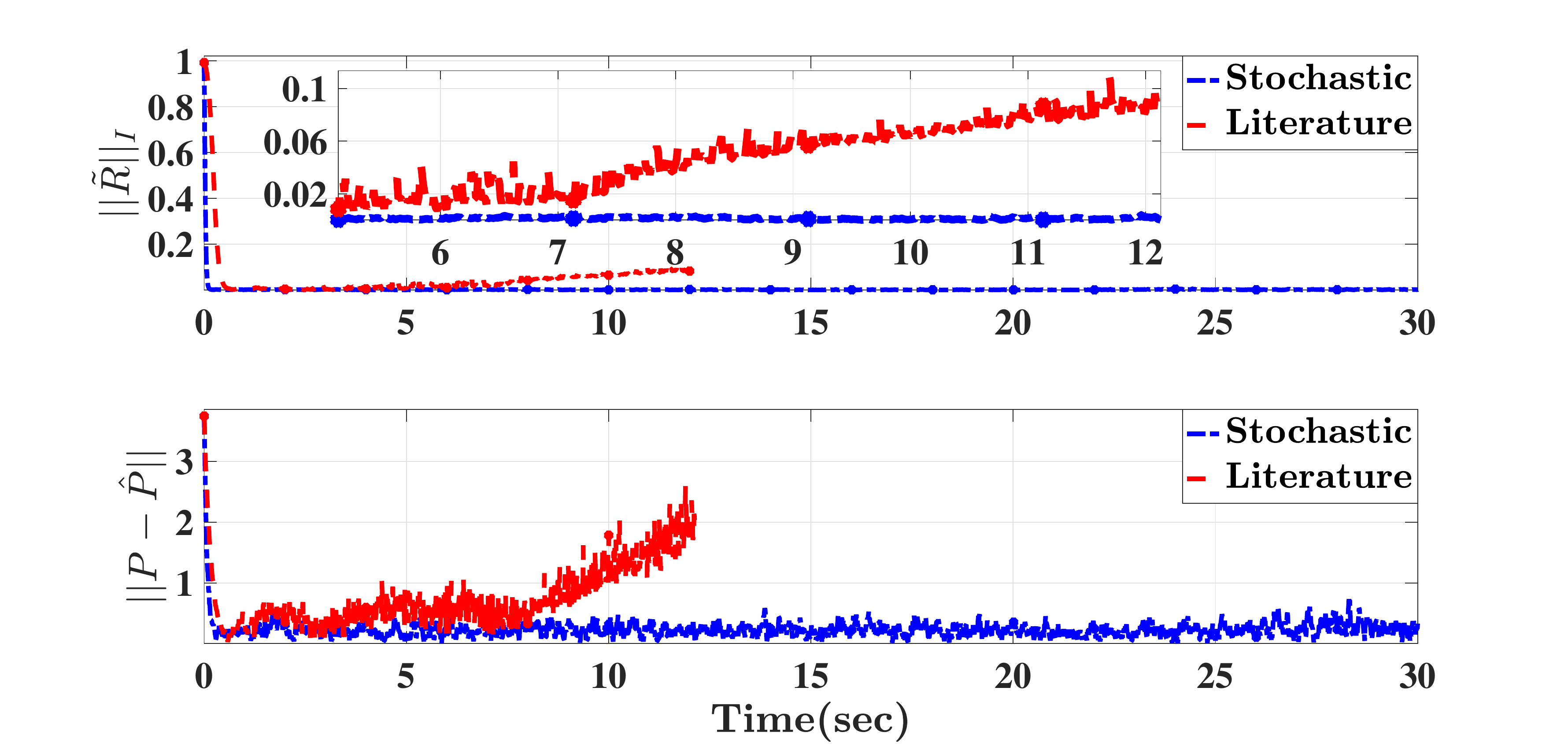}\caption{Tracking performance of normalized Euclidean distance error of $\left\Vert \tilde{R}\right\Vert _{I}$
		and Euclidean distance $\left\Vert P-\hat{P}\right\Vert $.}
	\label{fig:SE3STCH_7} 
\end{figure}

\newpage{}A comparison between the proposed stochastic observer in
Theorem \ref{thm:SE3STCH_2} and the deterministic pose observer in
\cite{hua2011observer} is presented in Figure \ref{fig:SE3STCH_7}.
The upper portion of Figure \ref{fig:SE3STCH_7} illustrates the normalized
Euclidean distance $\left\Vert \tilde{R}\right\Vert _{I}$, while
the lower portion presents the Euclidean distance $\left\Vert P-\hat{P}\right\Vert $
for both observers such that $\tilde{R}=\hat{R}R^{\top}$. Figure
\ref{fig:SE3STCH_7} shows stable output performance of the stochastic
observer with $\left\Vert \tilde{R}\right\Vert _{I}$ and $\left\Vert P-\hat{P}\right\Vert $
being regulated very close to the neighborhood of the origin confirming
the results shown in Figure \ref{fig:SE3STCH_5} and \ref{fig:SE3STCH_6}.
On the other side, the deterministic filter shows high oscillatory
performance before it goes out of stability.

\newpage{}Let $\mathring{{\rm v}}_{1}^{\mathcal{B}\left({\rm L}\right)}=R^{\top}\left({\rm v}_{1}^{\mathcal{I}\left({\rm L}\right)}-P\right)$
and $\mathring{{\rm v}}_{i}^{\mathcal{B}\left({\rm R}\right)}=R^{\top}{\rm v}_{i}^{\mathcal{I}\left({\rm R}\right)}$
denote the true body-frame vectors for $i=1,2,3$. Consider the error
between the true and measured body-frame vectors $\tilde{{\rm v}}_{1}^{\mathcal{B}\left({\rm L}\right)}={\rm v}_{1}^{\mathcal{B}\left({\rm L}\right)}-\mathring{{\rm v}}_{1}^{\mathcal{B}\left({\rm L}\right)}$
and $\tilde{{\rm v}}_{i}^{\mathcal{B}\left({\rm R}\right)}={\rm v}_{i}^{\mathcal{B}\left({\rm R}\right)}-\mathring{{\rm v}}_{i}^{\mathcal{B}\left({\rm R}\right)}$.
In the same spirit, let the error between the true and measured velocities
be given by $\tilde{\Omega}=\Omega_{m}-\Omega$ and $\tilde{V}=V_{m}-V$.
Table \ref{tab:SE3STCH_1} provides mean and STD of the input measurements
and the output data. It should be stressed that the mean errors of
$\left\Vert \tilde{R}\right\Vert _{I}$ and $P-\hat{P}$ approach
zero while the STD of $\left\Vert \tilde{R}\right\Vert _{I}$ is less
than its mean, and the STD of $P-\hat{P}\approx0.1$. Numerical results
outlined in Table \ref{tab:SE3STCH_1} affirm the robustness of the
proposed nonlinear stochastic filter as demonstrated in Figure \ref{fig:SE3STCH_5},
\ref{fig:SE3STCH_6}, and \ref{fig:SE3STCH_7}.

\begin{table}[H]
	\caption{Statistical analysis of the noise and bias in input measurements and
		output data of the proposed filter.}
	\begin{centering}
		\begin{tabular}{l|c|c|c|c|c}
			\hline 
			\multicolumn{6}{c}{Input measurements}\tabularnewline
			\hline 
			\hline 
			Index  & $\tilde{{\rm v}}_{1}^{\mathcal{B}\left({\rm L}\right)}$  & $\tilde{{\rm v}}_{1}^{\mathcal{B}\left({\rm R}\right)}$  & $\tilde{{\rm v}}_{2}^{\mathcal{B}\left({\rm R}\right)}$  & $\tilde{\Omega}\left({\rm rad/sec}\right)$  & $\tilde{V}\left({\rm m/sec}\right)$\tabularnewline
			\hline 
			\multicolumn{1}{c|}{Mean} & $\left[\begin{array}{c}
			0.15\\
			0.1\\
			-0.1
			\end{array}\right]$  & $\left[\begin{array}{c}
			-0.1\\
			0.1\\
			0.05
			\end{array}\right]$  & $\left[\begin{array}{c}
			0\\
			0\\
			0.1
			\end{array}\right]$  & $\left[\begin{array}{c}
			0.1\\
			-0.1\\
			0.1
			\end{array}\right]$  & $\left[\begin{array}{c}
			0.2\\
			0.5\\
			0.1
			\end{array}\right]$\tabularnewline
			\hline 
			STD  & $0.1\times\underline{\mathbf{1}}_{3}$  & $0.1\times\underline{\mathbf{1}}_{3}$  & $0.1\times\underline{\mathbf{1}}_{3}$  & $0.15\times\underline{\mathbf{1}}_{3}$  & $0.15\times\underline{\mathbf{1}}_{3}$\tabularnewline
			\hline 
			\hline 
			\multicolumn{6}{c}{Output data over the period (1-30 sec)}\tabularnewline
			\hline 
			\hline 
			Index  & \multicolumn{3}{c|}{$\left\Vert \tilde{R}\right\Vert _{I}$} & \multicolumn{2}{c}{$P-\hat{P}\left({\rm m}\right)$}\tabularnewline
			\hline 
			Mean  & \multicolumn{3}{c|}{$1.2\times10^{-3}$} & \multicolumn{2}{c}{$\left[-17.7,2.6,-8.4\right]^{\top}\times10^{-3}$}\tabularnewline
			\hline 
			STD  & \multicolumn{3}{c|}{$8.5\times10^{-4}$} & \multicolumn{2}{c}{$\left[1.15,1.07,1.27\right]^{\top}\times10^{-1}$}\tabularnewline
			\hline 
		\end{tabular}
		\par\end{centering}
	\label{tab:SE3STCH_1} 
\end{table}

Simulations presented in this section demonstrate the robustness of
the proposed stochastic filter in the sense of Stratonovich against
high levels of bias and noise components introduced in angular velocity,
translational velocity and vectorial measurements. Also, they show
that the stochastic filter is capable of correcting its position and
attitude even in presence of large initial error in a small amount
of time. In addition, the stochastic filter is autonomous, and therefore
no prior information about the upper bound of the covariance matrix
$\mathcal{Q}^{2}$ is required to achieve impressive estimation performance.

\section{Conclusion \label{sec:SE3STCH_Conclusion}}

Pose is naturally nonlinear and is modeled on the Special Euclidean
Group $\mathbb{SE}\left(3\right)$. Pose estimators used to be designed
as nonlinear deterministic filters neglecting the noise inherent to
the model dynamics. This is reflected in the nonlinear deterministic
filter design as well as in the potential function selection. In this
work, the pose problem has been formulated as a nonlinear pose problem
on $\mathbb{SE}\left(3\right)$. The problem is mapped from $\mathbb{SE}\left(3\right)$
to vector form using Rodriguez vector parameterization and position.
The problem is defined stochastically in the sense of Stratonovich.
Next, a nonlinear stochastic pose filter on $\mathbb{SE}\left(3\right)$
has been proposed. It has been shown that errors in position, Rodriguez
vector and estimates are semi-globally uniformly ultimately bounded
(SGUUB) in mean square and that they converge to the small neighborhood
of the origin for the case when noise is attached to the pose dynamics.
Simulation results prove fast convergence from large initialized pose
error even when angular and translational velocity vectors as well
as body-frame measurements are subject to high levels of noise and
bias.

\section*{Acknowledgment}

The authors would like to thank University of Western Ontario for
the funding that made this research possible. Also, the authors would
like to thank \textbf{Maria Shaposhnikova} for proofreading the article.

\section*{Appendix A \label{sec:SE3STCH_AppendixB}}
\begin{center}
	\textbf{\large{}{}{}An Overview on SVD in }{\large{}{}{}\cite{markley1988attitude}} 
	\par\end{center}

Let $R\in\mathbb{SO}\left(3\right)$ be the true attitude. The attitude
can be reconstructed through a set of vectors given in \eqref{eq:SE3STCH_Vect_R}.
Let $s_{i}$ be the confidence level of measurement $i$ such that
for $n$ measurements we have $\sum_{i=1}^{n}s_{i}=1$. In that case,
the corrupted reconstructed attitude $R_{y}$ can be obtained by 
\[
\begin{cases}
\mathcal{J}\left(R\right) & =1-\sum_{i=1}^{n}s_{i}\left(\upsilon_{i}^{\mathcal{B}}\right)^{\top}R^{\top}\upsilon_{i}^{\mathcal{I}}\\
& =1-{\rm Tr}\left\{ R^{\top}B^{\top}\right\} \\
B & =\sum_{i=1}^{n}s_{i}\upsilon_{i}^{\mathcal{B}}\left(\upsilon_{i}^{\mathcal{I}}\right)^{\top}=USV^{\top}\\
U_{+} & =U\left[\begin{array}{ccc}
1 & 0 & 0\\
0 & 1 & 0\\
0 & 0 & {\rm det}\left(U\right)
\end{array}\right]\\
V_{+} & =V\left[\begin{array}{ccc}
1 & 0 & 0\\
0 & 1 & 0\\
0 & 0 & {\rm det}\left(V\right)
\end{array}\right]\\
R_{y} & =V_{+}U_{+}^{\top}
\end{cases}
\]
For more details visit \cite{markley1988attitude}.

\bibliographystyle{IEEEtran}
\bibliography{arXiv_Source_Frank_SE3}

 \section*{AUTHOR INFORMATION}
 \vspace{10pt}
 {\bf Hashim A. Hashim} is a Ph.D. candidate and a Teaching and Research Assistant in Robotics and Control, Department of Electrical and Computer Engineering at the University of Western Ontario, ON, Canada.\\
 His current research interests include stochastic and deterministic filters on SO(3) and SE(3), control of multi-agent systems, control applications and optimization techniques.\\
 \underline{Contact Information}: \href{mailto:hmoham33@uwo.ca}{hmoham33@uwo.ca}.
 \vspace{20pt}
 
 {\bf Lyndon J. Brown} received the B.Sc. degree from the U. of Waterloo, Canada in 1988 and the M.Sc. and PhD. degrees from the University of Illinois, Urbana-Champaign in 1991 and 1996, respectively. He is an associate professor in the department of electrical and computer engineering at Western University, Canada. He worked in industry for Honeywell Aerospace Canada and E.I. DuPont de Nemours.\\
 His current research includes the identification and control of predictable signals, biological control systems, welding control systems and attitude estimation.
 \vspace{20pt}
 
 {\bf Kenneth McIsaac} received the B.Sc. degree from the University of Waterloo, Canada, in 1996, and the M.Sc. and Ph.D. degrees from the University of Pennsylvania, in 1998 and 2001, respectively. He is currently an Associate Professor and the Chair of Electrical and Computer Engineering with Western University, ON, Canada. \\
 His current research interests include computer vision and signal processing, mostly in the context of using machine intelligence in robotics and assistive systems.

\end{document}